%% file: GlasgowPreprint200811_arxiv.tex
\documentclass[12pt]{article}
\usepackage[dvips]{color}
\usepackage{epsfig}           % EPS manipulation with \epsfig
\usepackage{graphics}
\usepackage{color}
\usepackage{colortbl}
\usepackage{glas}

\begin{document}
%%begin new command to define color in table njaka 20-03-08
\renewcommand*{\DefineNamedColor}[4]{%
    \textcolor[named]{#2}{\rule{7mm}{7mm}}\quad
    \texttt{#2}\strut\\}
%%end new command
% ************************************************************************
% Glasgow template !!!
\begin{titlepage}{GLAS-PPE/2008-11}{8$^{\underline{\rm{th}}}$ August 2008}
\title{Characterisation of the VELO High Voltage System}%
\author{Barinjaka Rakotomiaramanana$^1$, Chris Parkes$^1$, Lars Eklund $^1$\\
\\
$^1$ University of Glasgow, Glasgow G12 8QQ, Scotland\\}
%$^2$ Rutherford Appleton Laboratory, Didcot
\vspace*{1.0cm}
\begin{abstract}
The high voltage system supplies the bias voltage to the 88 silicon sensors
which comprise the LHCb Vertex Locator (VELO). This note describes the results of the tests which have been performed on the hardware of the high voltage system of the VELO. Each individual test detailed in this note corresponds to a specific requirement of the system. These requirements arise primarily from ensuring the safety of the silicon sensors and the quality of the data taken from the VELO modules.
The tests performed are in four categories: normal operation of the high voltage system; verification of its stability under operation; discussion of its behaviour in failure modes; and details of operation at low voltage.
Noteworthy issues, identified through the tests, include  the behaviour of the high voltage modules at voltages below 9V, the current limit that can be applied during ramping of the voltage, and the speed with which the voltage is cut during failures of the system. The results of these tests provide high confidence that the high voltage system can be safely used and demonstrate that low noise is produced by the power supplies.
\end{abstract}
\vspace*{1.0cm}
\begin{center}
\textit{LHCb Public Note, LHCb-2008-009}
\end{center}
\newpage
\end{titlepage}
% ************************************************************************
%\sffamily

%-------------------------------------------------------------------------
%     Title + Abstract
%-------------------------------------------------------------------------
%\include{lhcbnote_title}

%-------------------------------------------------------------------------
%     Table of Contents / List of Tables / List of Figures
%-------------------------------------------------------------------------
\tableofcontents 
\newpage  
\addvspace{10pt} 
\listoftables 
\newpage 
\addvspace{10pt} 
\listoffigures 
\newpage 

%-------------------------------------------------------------------------
%     Main Text
%-------------------------------------------------------------------------
%\include{lhcbnote_text}
\input{introduction/introduction.tex}

\input{system_requirements/requirements.tex}
\input{normal_operation/normal_operation.tex}

\input{stability_of_operation/stability_of_operation.tex}

\input{failure_modes/failure_modes.tex}

\input{cables_and_patch_panels_testing/cables_and_patch_panels_testing.tex}
\input{low_voltage_test/low_voltage_test.tex}
\input{summary/summary.tex}

\input{conclusion/conclusion.tex}

%-------------------------------------------------------------------------
%     Appendix
%-------------------------------------------------------------------------
%\appendix
%\include{lhcbnote_appendix}

%-------------------------------------------------------------------------
%     Bibliography
%-------------------------------------------------------------------------
%\include{lhcbnote_biblio}

%-------------------------------------------------------------------------
%     Tables
%-------------------------------------------------------------------------
%\include{lhcbnote_tables}

%-------------------------------------------------------------------------
%     Figures
%--------------------------------------------------------------------------
%\include{lhcbnote_figures}
%
%
\end{document}

%% file: introduction/introduction.tex
\section{Introduction}

The high voltage system supplies the bias voltage to the 88 silicon sensors
which comprise the LHCb Vertex Locator (VELO). The main hardware components of the high voltage system are: the power supply crate; the high voltage modules; the counting house patch panels; the detector patch panels; and all connecting cables. A detailed description of the hardware system is available in \cite{HVC}. The software of the high voltage system will be described in a future note.

This note describes the tests which were performed to the high voltage system hardware. The structure of this document is as follows. After this introduction, Section \ref{requirements} gives the requirements of the high voltage system. Section \ref{normal_operation} describes the tests of the high voltage modules under normal operation such as voltage ramping, and tests of the software and hardware limits.
The tests dedicated to demonstrating the stability of the high voltage modules with time and assessing the noise of the system are detailed in Section \ref{stability_of_operation}. Section \ref{failure_modes} shows the behaviour of the high voltage modules and crate under failure modes and includes a description of the two possible modes of operation of the high voltage module: current control mode, and current trip mode. The tests performed to the cables and patch panels are detailed in Section \ref{cables_and_patch_panels_testing}. Section \ref{low_voltage_test} describes 
the tests which were performed at low voltages. Finally, in Section \ref{summary} 
all the tests performed on the nine high voltage modules are summarized in a table. This table details which individual channels have had each test applied. Conclusions are provided in Section \ref{conclusions}.

%% file: system_requirements/requirements.tex
\section{System requirements}\label{requirements}

The high voltage system has been designed to comply with the requirements of the VELO module and the constraints imposed by operating the VELO in an extreme radiation environment.
\noindent The high voltage system requirements are as follows:
\begin{itemize}
\item Supply sufficient voltage to fully deplete the VELO when it is
irradiated to the expected dose and in the expected annealing scenario. The VELO system is specified for 500V operation at a few mA. The high voltage power supply module chosen supplies up to 700V at 4mA. All components of the system have been tested to 500V.

\item Ensure the safety of the VELO modules and its operators. The system implements a current trip limit in both hardware and software. The operation of the system can also be inhibited through an interlock system \cite{interlock}.

\item Provide a low noise level in order to maintain the high signal to noise ratio obtained from the VELO silicon modules.

\item Protect the VELO against high voltage discharges in the vacuum. The electronics hybrid of the VELO sensors implements guard traces around the high voltage bias line to prevent discharges. The high voltage system also provides the voltage for these guard traces.

\item The system must be remotely controllable and integrated in to the LHCb control system. No components requiring regular maintenance can be in the radiation zone of the experiment.
\end{itemize}

%% file: normal_operation/normal_operation.tex
\section{Normal operation}\label{normal_operation}

This section describes the tests performed to verify the operation of the high voltage module under standard operating conditions. The tests verify that the voltage can be set correctly and can be ramped up or down on every channel. 
 
\input{normal_operation/part_1.tex}

%% file: normal_operation/part_1.tex
\subsection{Voltage ramping up and ramping down}\label{label_one}
\begin{figure}[!ht]
\epsfig{file=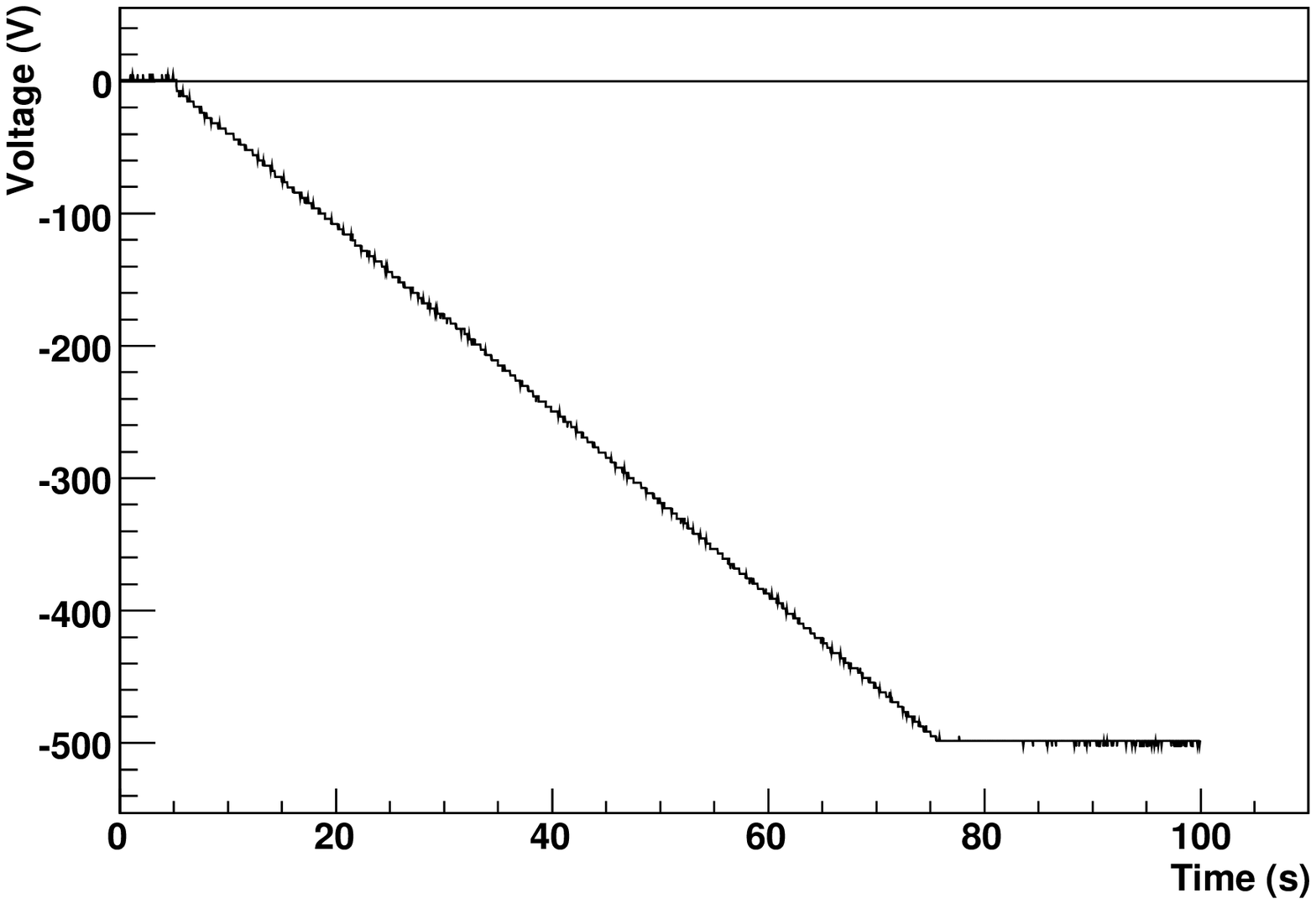,width=0.5\textwidth}
\epsfig{file=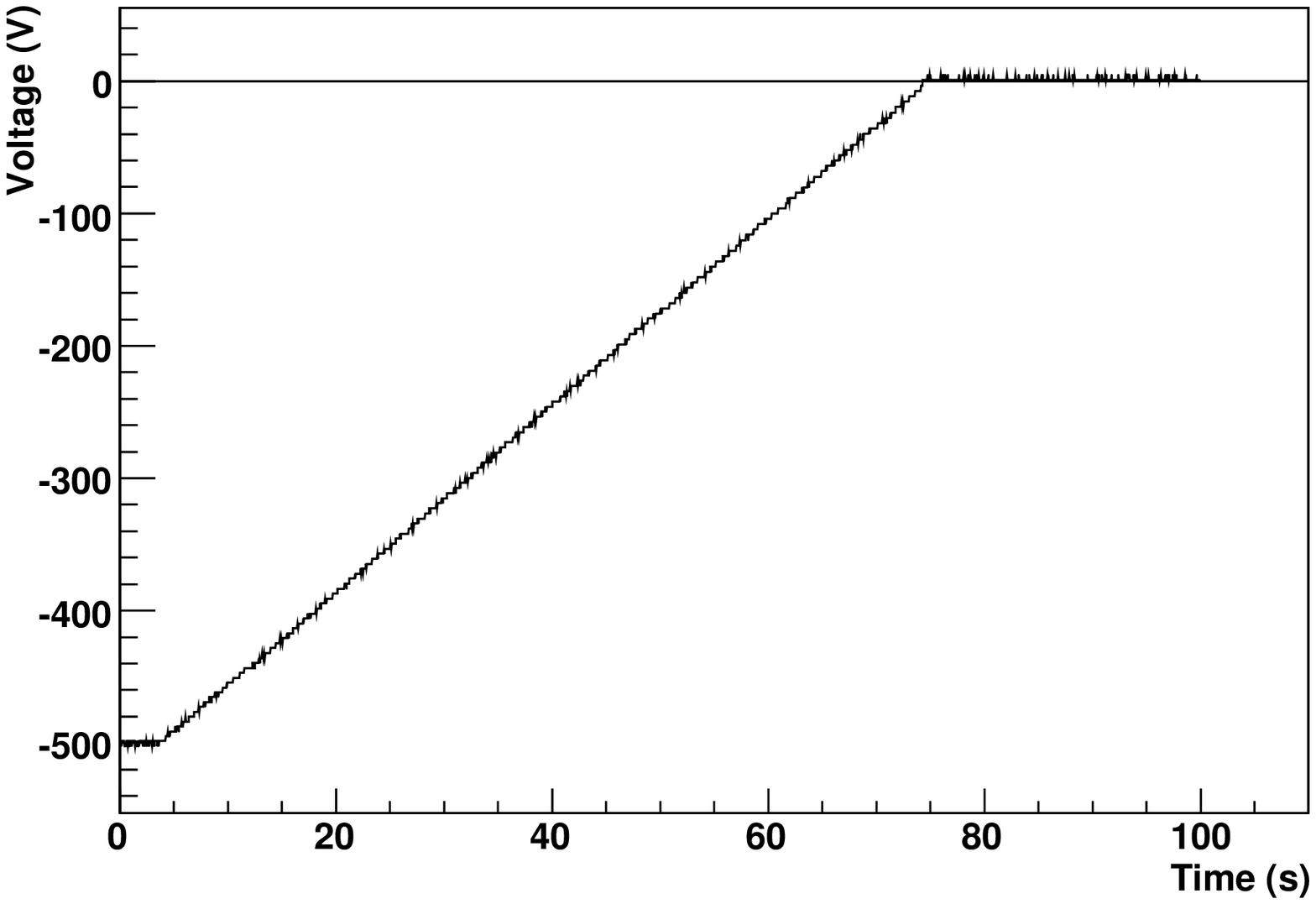,width=0.5\textwidth}
\caption[Voltage measured on an oscilloscope during ramping] {Voltage measured on an oscilloscope during ramping of channel 15 of the high voltage module with serial number 474185 07/2006. The performance when ramping from 0V to -500V (left) and from -500V to 0V (right) is shown.}
\label{ramping}
\end{figure}

The first test verifies that the full range of voltage can be applied to each channel of each high voltage module. Initially the voltage was ramped from 0V to -500V and then from -500V back to 0V. The control was performed through software and a ramping speed of 7V/s was used. The voltage produced by the high voltage module was measured as a function of time with an oscilloscope. Two plots were produced for each channel: one for ramping up and
one for ramping down, see Figure \ref{ramping}. These two plots were produced for each channel of all 9 high voltage modules. No problems were observed.

\subsection{Comparison of channel ramping up and down gradients}\label{label_two}
\begin{figure}[!ht]
\epsfig{file=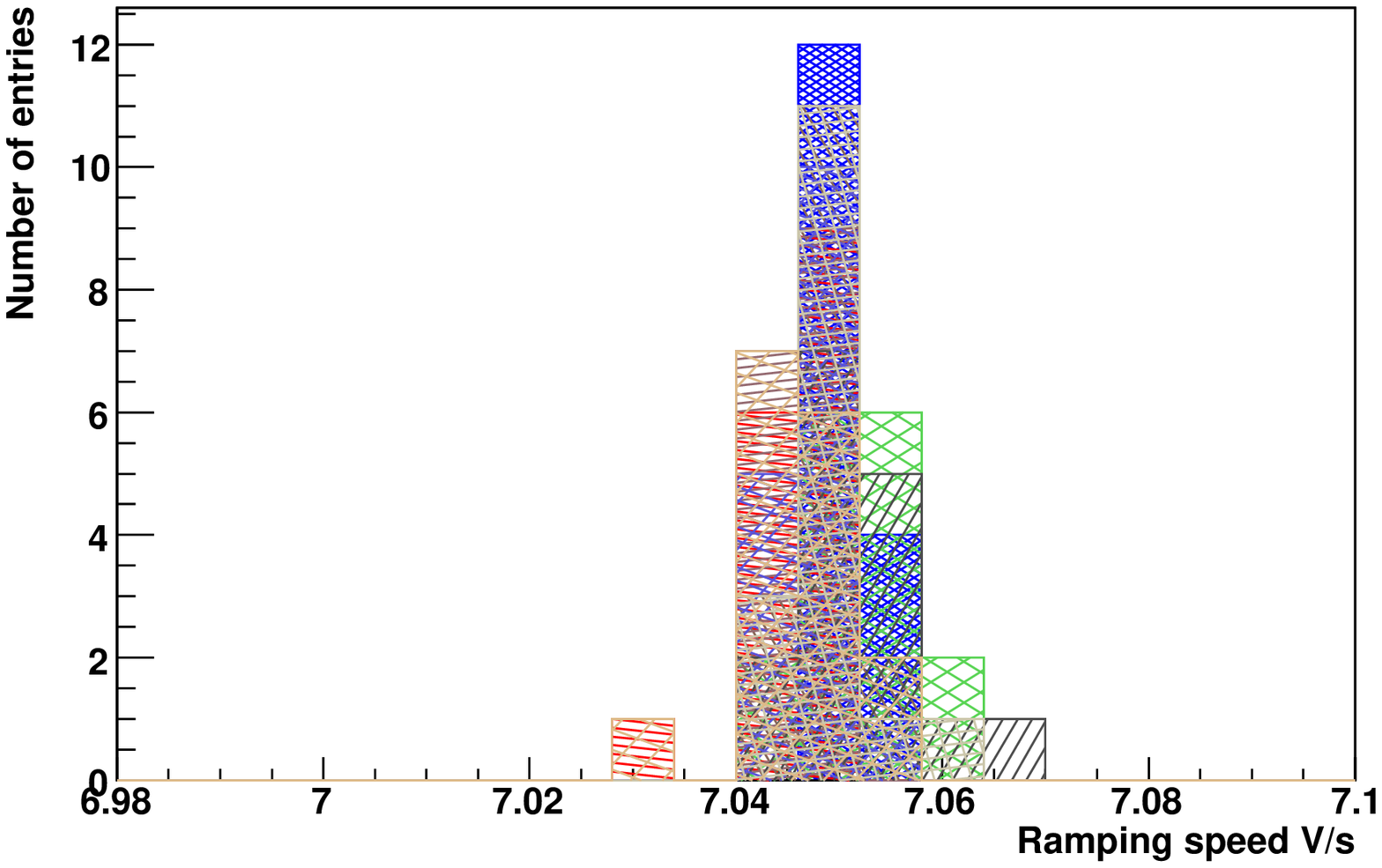,width=0.5\textwidth}
\epsfig{file=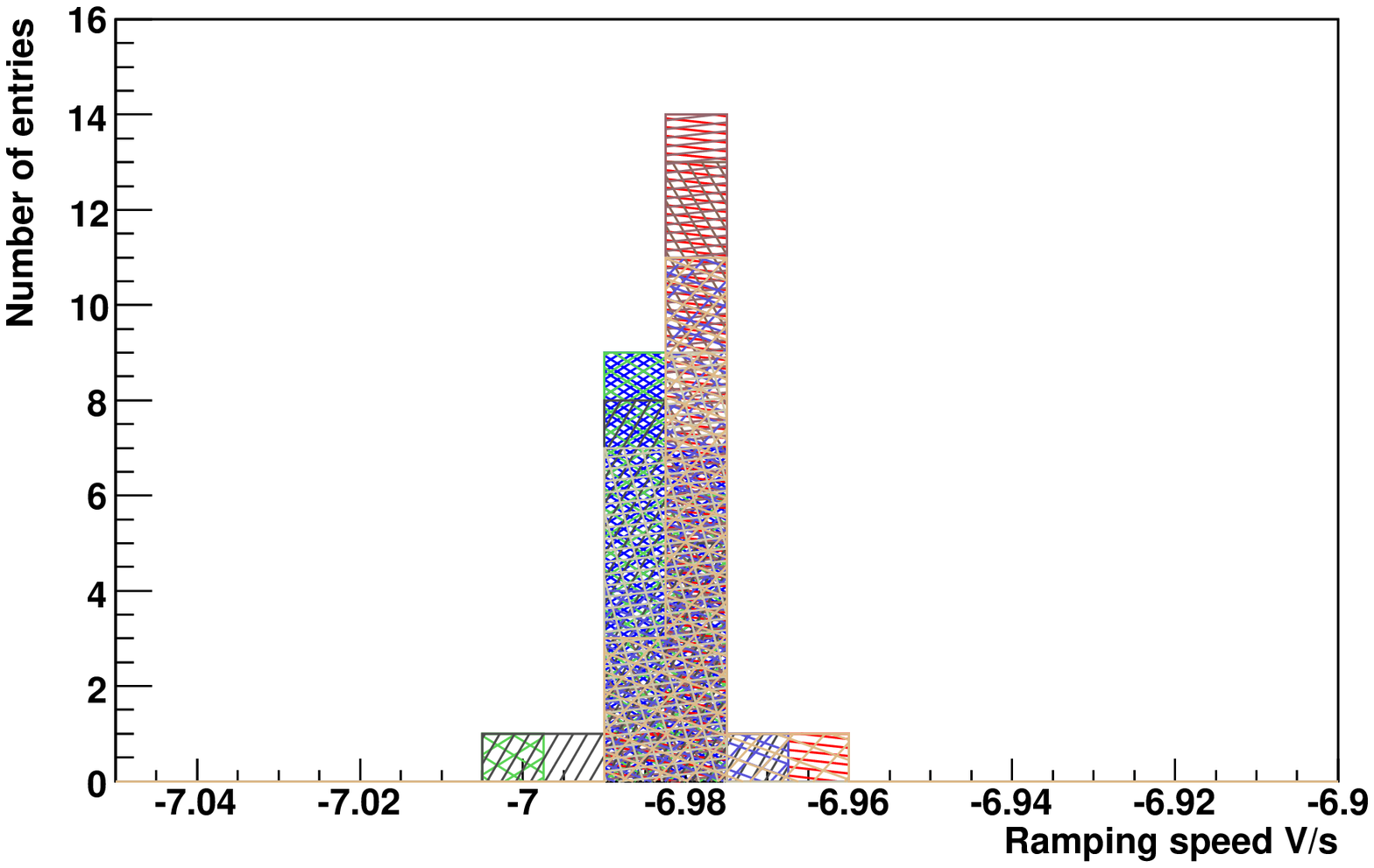,width=0.5\textwidth}
\caption[Variation of the ramping speeds measured on each individual module by applying voltage from 0V to -500V and back to 0V.]{Variation of the ramping speeds
 measured on each individual module by applying voltage (left) from 0V to -500V and (right) from -500V to 0V. 
One colour represents the group of 16 channels of one high voltage module, all 9 modules are shown but are not clearly visible as the plots are superimposed.}  
\label{ramping_graphs}
\end{figure}

Using the data from the previous test, the consistency of the ramping speeds in each channel of each module can be tested. Each power supply module has 16 channels. The gradient of ramping, up and down, is calculated for all 16 channels of each module. The nominal ramping speed set though the software is 7V/s. The results of this check are shown in Figure \ref{ramping_graphs}. The ramping speeds calculated during ramping up from 0V to 500V are slightly higher than the ramping speed set through software, while those for ramping down back to 0V are mostly slightly less than the ramping speed set through software. The variance of the ramping speeds from channels of different modules is larger than that inside a single modules. All channels of all modules performed acceptably and were within 1\% of the set value.

\subsection{Firmware error for ramping down}\label{firmware_error}

The stress test that were performed on the Iseg Modules have revealed a bug in the firmware that is currently installed in the modules. At the time of writing, the firmware version used is 2.04 for the modules. This error can cause a channel to instantaneously change its voltage, rather than ramping at the required speed. This could potentially damage the VELO module.

The error occurs if one (or more) channel(s) is being ramped and another channel is turned on. In this case the first channel(s) instantaneously jumps to its/their target voltage when the second channel is turned on, rather than continuing to ramp. The problem occurs for any channels in the same board of a power supply module: there are two boards in each power supply module, one for the first eight channels and one for the second eight. The problem occurs as a result of issuing the command to turn on a channel: i.e. it occurs even if that channel is already turned on. The problem only occurs if the channels are being ramped towards 0V (e.g. from -500V to -400V), not if the channels are being ramped away from 0V (e.g. from -400V to -500V).

The problem is caused by a firmware error and hence occurs whatever the method used to control the modules (standalone Iseg software or PVSS).
The problem has been reported and was reproduced by Iseg. Iseg has released a new patch for the high voltage module
firmware, version e08f0\_222.hex, which can be flashed into the FPGAs using the version 1.58 of the IsegHVControl. However, as we have previously had a bad experience in upgrading the firmware of a spare module (making the module unusable without return to the company), and the time it would take to repeat all necessary testing, we will not upgrade for 2008 operation.

Instead of performing the firmware update, we have introduced a protection in the PVSS FSM. This protection prevents the issuing of a second recipe that could cause this problem, while a first recipe is still producing a ramping . The channel requested to turn on will not be turned on until after all channels on that power supply module board have finished ramping. In practice, this means the system does not immediately respond to the second recipe command, but performs this action only later after all channels from the first recipe command have finished ramping. The protection is there for ramping in either voltage direction, due to the practicalities of its implementation. This FSM protection only protects for recipes, it is still possible to cause this problem by controlling individual channels. Control of individual channels should only be performed by VELO HV experts.

%% file: stability_of_operation/stability_of_operation.tex
\section{Stability of operation}\label{stability_of_operation}

This section reports on tests to ensure that the output of the module is stable over time and does not have voltage spikes, and to measure the noise level at the output of the power supply and in the final setup. Significant voltage spikes could potentially damage the sensors, or at lower values would contribute to the noise thus degrading the performance of the VELO.

\subsection{Voltage spikes}

Two methods were used to check for any voltage spikes in the output of the power supply. The methods were applied both for constant output voltages over significant times and when ramping the voltage on the power supply.

\subsubsection{First method for static operation - Oscilloscope in trigger mode}\label{label_eight}

The output of a channel was connected to an oscilloscope through a
voltage divider which attenuates the input voltage by a factor of 20.
The oscilloscope was set to trigger if a voltage spike occurred. 

\noindent Each test was performed under the following conditions:
\begin{itemize}
\item The voltage applied was -100V which corresponds to -5V displayed on the oscilloscope.
\item The oscilloscope was put in the single trigger mode.
\item The trigger level was set at -5.12V so that if there was a voltage above -102.4V the trace would be saved during the acquisition time.
\item The oscilloscope was left armed for at least 12 hours.
\end{itemize}

No spikes were observed during any of the tests.

\subsubsection{Second method for static operation - Oscilloscope in persistency mode}\label{label_nine}

The oscilloscope was also use to test for voltage spikes using AC coupling and in persistency mode.

\noindent The power supply channel output was connected directly to the oscilloscope and the tests were performed under the following conditions:

\begin{itemize}
\item Voltage applied was -140 V.
\item The oscilloscope was in persistency mode, AC coupling was used, and the trigger level was set at 52mV.
\item The operation was performed for 12 hours.
\end{itemize}

\begin{figure}[!ht]
\epsfig{file=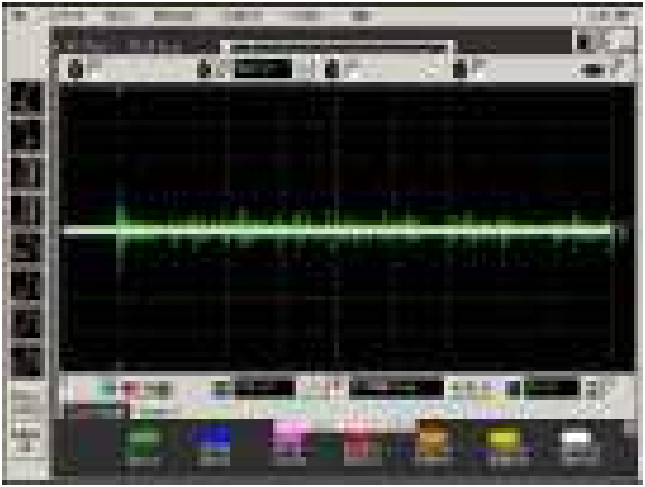,width=0.50\textwidth}
\caption[Oscilloscope trace showing a test for voltage spikes] {Oscilloscope trace showing a test for voltage spikes. The trace was taken in persistency mode for 12 hours with -140V applied. The voltage scale is 500mV/div. The time scale is 500ns/div.}
\label{picture_four}
\end{figure}

Figure \ref{picture_four} shows that the biggest spike observed is around 500mV.

\subsubsection{First method during ramping - Oscilloscope in trigger mode}\label{label_ten}

In addition to checking for spikes during constant voltage operation, it is important to check that no significant spikes are obtained when turning the bias voltage on or off. Hence the previous tests were repeated while ramping the voltage.
\noindent The power supply channel output was connected directly to the oscilloscope and the tests were performed under the following conditions:

\begin{itemize}
\item The system was ramped from 0V to 60V, and repeated from 60V to 120V.
\item The oscilloscope was put in DC mode. 
\item The trigger level was adjusted to find the minimum value such that a trace was not acquired.
\end{itemize}

The minimum value which did not provoke a trace acquisition was $Vset$ + $0.9V$. Hence, we conclude that the power supply overshoots the set voltage and does not produce spikes at more than a 1V level. 

\subsubsection{Second method during ramping - Oscilloscope in persistency mode}\label{label_eleven}

A check for voltage spikes was also performed using the oscilloscope in persistency mode.

\noindent The power supply channel output was connected directly to the oscilloscope and the tests were performed under the following conditions:

\begin{itemize}
\item The system was ramped in voltage from 0 to -200V.
\item The oscilloscope was in persistency mode, AC coupling was used, and the trigger level was set at 100mV.
\end{itemize}

\begin{figure}[!ht]
\epsfig{file=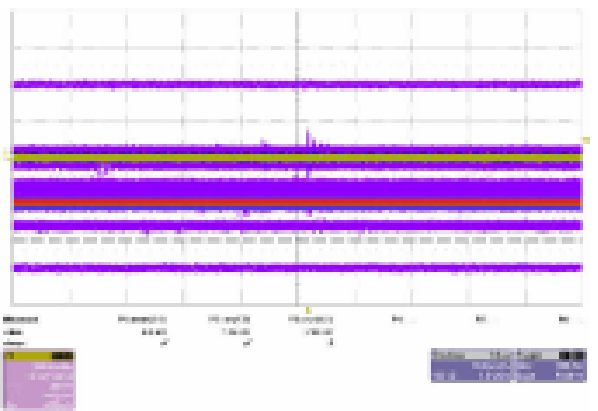,width=0.50\textwidth}
\epsfig{file=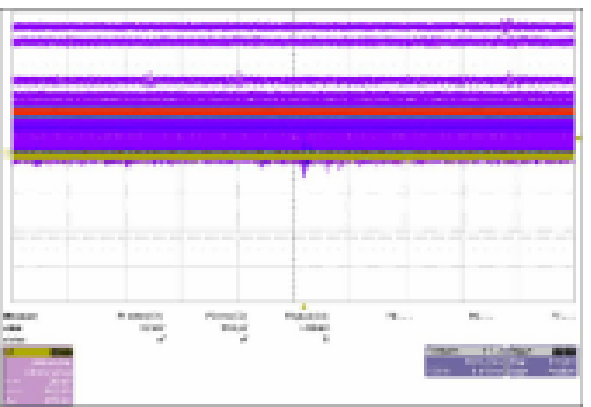,width=0.50\textwidth}
\caption[Oscilloscope traces showing tests for voltage spikes during ramping of the voltage]{Oscilloscope traces showing a test for voltage spikes during ramping of the voltage. The trace was taken in persistency mode. A- (left) The voltage was ramped from 0V to -200V. B- (right) The voltage was ramped from -200V to 0V. The voltage scale is 200mV/div. The time scale is 10$\mu$s/div.}
\label{picture_three}
\end{figure}

Figure \ref{picture_three}A shows the oscilloscope trace when the voltage ramps up from 0V to -200V and
Figure \ref{picture_three}B when it ramps down from -200V to 0V in persistency mode.
The two figures show that there are no obvious significant spikes during ramping.

\subsection{Noise measurement}

The tests in this section are intended to characterise the noise of the power supply module and of the full system. The first two tests characterise the noise and potential common mode noise in the power supply. The last two tests use the full assembled system and include a measurement of the bias voltage noise with a dummy load equivalent to connecting a module.

\subsubsection{Noise at the module output}\label{label_twelve}

The oscilloscope was used to measure the power supply module noise.

\noindent The power supply channel output was connected directly to the oscilloscope and the tests were performed under the following conditions:
\begin{itemize}
\item The Voltage applied was -100V.
\item The oscilloscope was used to measure the peak to peak voltage, RMS voltage and determine the frequency spectrum of the noise at the same time. The frequency spectrum was obtained by performing a Fast Fourier Transform of the output.
\end{itemize}

\begin{figure}[!ht]
\epsfig{file=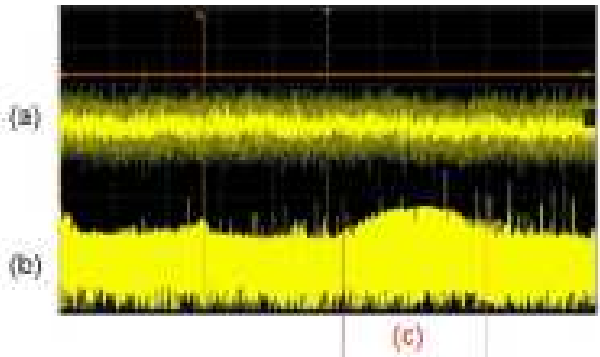,width=0.50\textwidth}
\caption[Oscilloscope trace of the noise spectrum at the module output]{Oscilloscope trace of the noise spectrum at the module output. (a) Indicates the noise level scan, the voltage scale is 100mV/div. (b) Shows the Fast Fourier Transform scan from 0Hz to 100MHz with (c) indicating a frequency range in which a slight noise excess is observed. }
\label{picture_one}
\end{figure}

Figure \ref{picture_one} shows the oscilloscope traces obtained. The following results were obtained:
\begin{itemize}
\item The Peak to peak voltage is approximately 100mV.
\item The RMS voltage is approximately 5mV.
\item The excess noise indicated by (c) in Figure \ref{picture_one} from the Fast Fourier Transform is in the frequency range of 50 MHz to 80 MHz.\end{itemize}

While the noise levels are relatively low the frequency range does correspond to that of the VELO system, since the LHC beam crossing and front-end chip clocking frequency is 40MHz.

\subsubsection{Common mode noise measurement}\label{label_thirteen}

\begin{figure}[!ht]
\epsfig{file=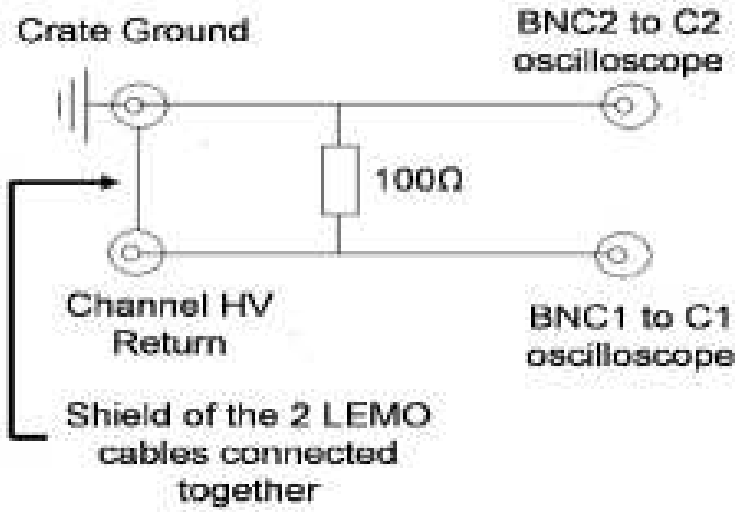,width=0.60\textwidth}
\caption[Schematic of the common mode noise measurement setup] {Schematic of the common mode noise measurement setup.}
\label{picture_first}
\end{figure}

A test was performed to measure the noise level produced by the potential difference between the ground of the power supply crate and the HV return of the channel. Figure \ref{picture_first} shows the schematic of the setup used to measure this common mode noise.

The channel HV return was connected directly to channel 1 of the oscilloscope and the ground of the crate to channel 2 of the same oscilloscope. The noise was displayed and measured on the oscilloscope by subtracting channel 2 from channel 1. 
\noindent The measurement on the oscilloscope was made in the following settings:
\begin{itemize}
\item Measurements have been performed at 0V and -140V on all channels.
\item The oscilloscope was used to measure the peak to peak voltage, RMS voltage and the frequency response by performing a Fast Fourier Transform at the same time.
\end{itemize}

\begin{figure}[!ht]
\epsfig{file=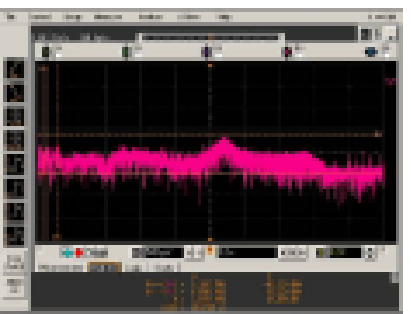,width=0.50\textwidth}
\epsfig{file=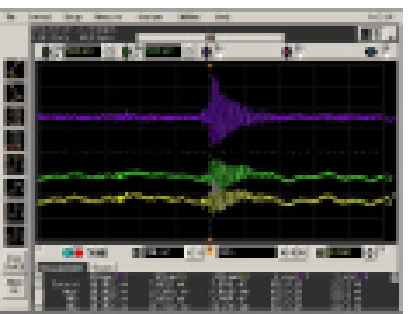,width=0.50\textwidth}
\caption[Oscilloscope traces of the common mode noise measurement] {Oscilloscope traces of the common mode noise measurement. A- (left) Fast Fourier Transform of the common mode noise. The frequency domain was from 0Hz to 100MHz. B- (right) The oscilloscope trace which shows the non Gaussian component of the noise. The time scale was 200 ns/div and voltage scale 4mV/div.}
\label{picture_eleven}
\end{figure}

Figure \ref{picture_eleven}A is an oscilloscope screen image showing an example noise distribution.  The following results were obtained:
\begin{itemize}
\item Peak to peak voltage approximately equal to 63mV.
\item RMS voltage approximately equal to 4.8mV.
\item The main excess noise peak indicated in Figure \ref{picture_eleven}A is at the frequency 54 MHz.
\item The common mode noise measured between the crate ground and the HV return contains a non Gaussian component which is shown in Figure \ref{picture_eleven}B. The `ringing' after the noise spike could be due to either the power supply module or the response of the oscilloscope. 

\subsubsection{Noise measurement in the LHCb pit without a filter}\label{label_fourteen}

A test was performed on the full high voltage system in the LHCb pit to measure the noise level at the end of the detector cable between the HV guard and the ground. This measurement was performed without any filter and hence corresponds to the noise that will be seen on the HV guard line under normal operation.

\noindent The guard line was connected directly to the oscilloscope and the tests were performed under the following conditions:
\begin{itemize}
\item The voltage applied was -100V.
\item The oscilloscope was used to measure the peak to peak voltage, RMS voltage and a Fast Fourier transform at the same time.
\end{itemize}
\begin{figure}[!ht]
\epsfig{file=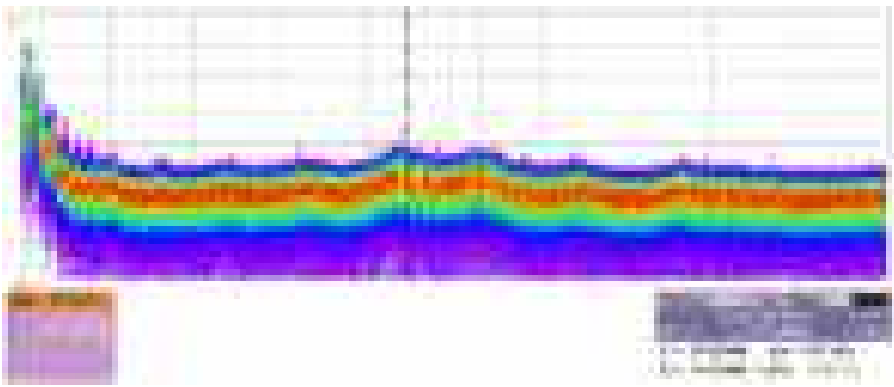,width=0.70\textwidth}
\caption[Oscilloscope trace showing the noise measured at the detector cable output between HV guard and ground.] {Oscilloscope trace showing the noise measured at the detector cable output between HV guard and ground. The time scale was 5$\mu$s/division. The frequency domain was from 0Hz to 100MHz. }
\label{picture_thirteen}
\end{figure}

Figure \ref{picture_thirteen} is a screen image from the oscilloscope. The following results were obtained:

\begin{itemize}
\item The Peak to peak voltage was approximately 11.3mV.
\item The RMS voltage was approximately 2.3mV.
\item The excess noise peaks indicated in Figure \ref{picture_thirteen} are at the frequencies of 44MHz and 53MHz.
\end{itemize}

\subsubsection{Noise measurement in the LHCb pit with a filter}\label{label_fifteen}

A measurement was performed of the noise level between the HV bias and ground in the LHCb pit under near realistic conditions. The hybrid of the VELO silicon module contains a low pass filter. A load emulating the hybrid was connected at the detector cable output. Figure \ref{picture_second} shows the schematic of the emulated hybrid. NOTE: this filter is not the same as that implemented on the hybrid and this test will be performed again.

\begin{figure}[!ht]
\epsfig{file=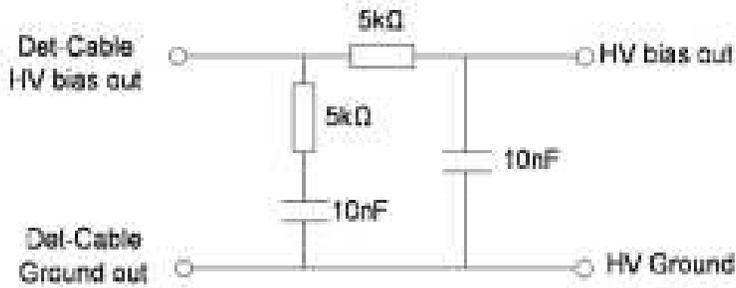,width=0.60\textwidth}
\caption[Schematic of the load emulating the hybrid with HV filter] {Schematic of the load emulating the hybrid with HV filter.}
\label{picture_second}
\end{figure}

\noindent The bias line was connected directly to the oscilloscope and the tests were performed under the following conditions:
\begin{itemize}
\item The Voltage applied was -100V.
\item The oscilloscope was used to measure the peak to peak voltage, RMS voltage and perform a Fast Fourier Transform in the same time.
\end{itemize}
\begin{figure}[!ht]
\epsfig{file=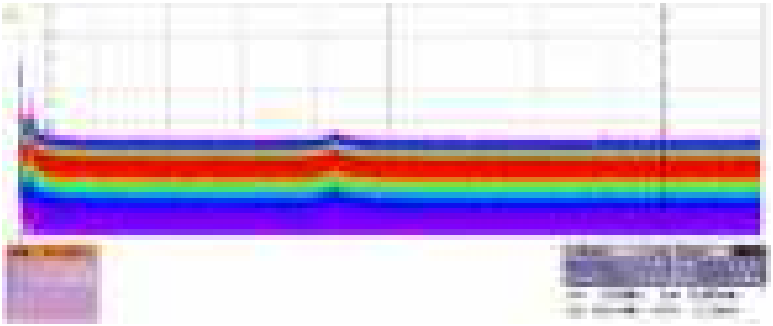,width=0.60\textwidth}
\caption[Oscilloscope trace of the noise measured at the detector cable output with HV filter] {Oscilloscope trace showing the frequency spectrum of the noise measured between HV bias and ground. The noise was measured at the output of an emulated hybrid which was connected to the detector cable.The frequency domain was from 0Hz to 100MHz.}
\label{picture_fourteen}
\end{figure}

Figure \ref{picture_fourteen} is a screen image from the oscilloscope. The following results were obtained:
\begin{itemize}
\item The peak to peak voltage was approximately 6.9mV.
\item The RMS voltage was approximately 1.2mV.
\item No significant excess noise was seen in the frequency spectrum (see Figure \ref{picture_fourteen}) 
\end{itemize}

%% file: failure_modes/failure_modes.tex
\section{Failure modes}\label{failure_modes}

The previous sections of this document have reported on the behaviour of the system under normal operation. This section concentrates on the behaviour under failure modes. The system response to a power failure or if the interlock signals are fired during operation is discussed. Tests of the current and voltage limits that can be set on the module (both in hardware and software) are reported. 

\input{failure_modes/power_cut}
\input{failure_modes/interlock.tex}
\input{failure_modes/currentlimit.tex}

%% file: failure_modes/power_cut.tex
\subsection{Power off}\label{label_sixteen}

A test was performed to understand the behaviour of the system under a power cut. The high voltage module crate was purchased equipped with an uninterruptible power supply (UPS) which should initially maintain power to the system and then ramp down the channels.

A crate containing two high voltage modules was used for this test and a voltage of -500V was applied to all channels. One of the channels was connected to the oscilloscope, through a voltage divider. The main power to the power supply crate was then turned off to simulate a power cut. 

\begin{figure}[!ht]
\epsfig{file=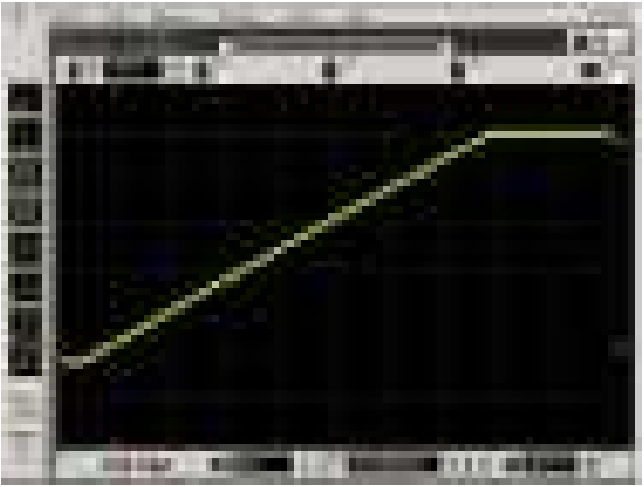,width=0.50\textwidth}
\caption[Oscilloscope trace of the output voltage when a power cut occurs.] {Oscilloscope trace shows the output voltage of the module when a power cut occurs. The time scale is 5s/division. }
\label{picture_fifteen}
\end{figure}

The voltage of -500V is maintained on the channels for approximately 10s, as seen in Figure \ref{picture_fifteen}A. It then takes approximately 40s to ramp down from -500V to 0V. After a further 10s the modules switch off completely. This behaviour of the UPS system provides a safer operation of the system than the sudden cut of voltage that would otherwise occur.

\subsection{Brown-out}\label{label_seventeen}

Another potential failure mode of the system is a temporary interruption of the power supply or 'Brown-out'. Measurements were made to determine the HV power supply module response with short interruptions.

A crate containing seven high voltage modules and with a voltage of -140V applied to all channels was used. One of the channels was connected directly to the oscilloscope with the following settings:
\begin{itemize}
\item The oscilloscope was in the single trigger mode and AC coupling.
\item Trigger level was set at 40mV.
\end{itemize}

\begin{figure}[!ht]
\epsfig{file=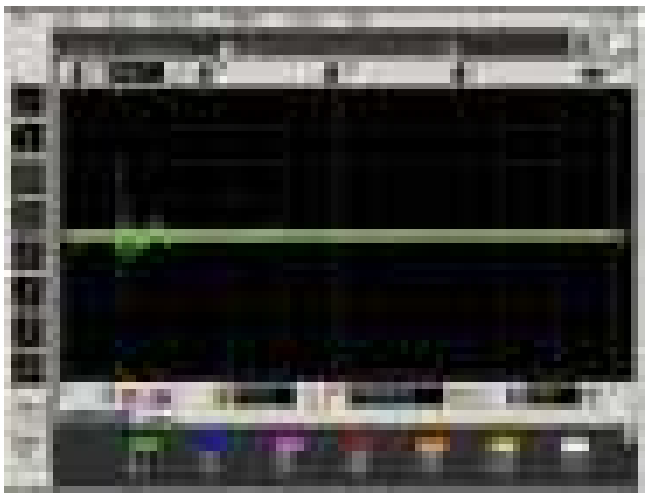,width=0.50\textwidth}
\caption[Oscilloscope trace of the output voltage during a brown-out.] {Oscilloscope trace of the output voltage during a brown-out when it is in persistency mode. The voltage scale is 500mV/division. The time scale is 500ns/division. }
\label{picture_sixteen}
\end{figure}

The main power to the crate was switched off and on again to simulate a brown-out of short duration, for example 2s. Figure \ref{picture_sixteen} shows that a brown-out causes a spike in the voltage observed, but that the highest spike observed was approximately 1V.

If the brown-out duration exceeds 10s the power supply module will already have started to ramp down the voltages on the channels. In this case, the channels continue to ramp to 0V and are turned off, as in the case of a longer term power cut.

%% file: failure_modes/interlock.tex
\subsection{Interlock Signals}\label{hv_enabled_off}

The VELO interlock system is a safety system that takes a number of inputs and will inhibit the operation of the VELO if a problem occurs, e.g. if a temperature alarm occurs from the temperature sensors on the modules. The interlock system provides an input signal to the power supply modules, if these signals are fired the power supply operation should be inhibited. A separate interlock signal is required for every channel of the power supply module. The measurements made in this section test the behaviour of the system when the interlock signals are fired.

In order to test the interlock system for the high voltage modules a box was constructed which supplies the required DC voltage of 5V. The box has 8 switches allowing control of the individual interlock signal to a set of 8 channels of one module.

\subsubsection{Interlock signals fired}\label{label_eighteen}

The response of the system was investigated when all the interlock signals are fired.

A crate containing 9 high voltage modules with the voltage set to -500V on all channels was used for this test. When all the interlock signals are fired the voltages are shut down, without ramping, and an inhibit error alarm occurs via software. Applying the interlock signal again to all channels and clearing the error alarm via the PVSS software, the inhibit behaviour was reset and all channels were operational again and could be ramped up on demand as required.

In addition it is important to verify that the interlock works for each individual channel of the system. Using the interlock test box the check was performed as follows:
\begin{itemize}
\item A voltage of -500V was applied to all channels.
\item HV interlock signals were removed one by one to individual channels.
\item The corresponding channel was checked to see that the voltage was interrupted.
\item Checks were performed to verify all the other channels in the module were still on and had -500V measured.
\end{itemize}

This test has shown that each channel can be inhibited individually without affecting the remaining channels.

\subsubsection{Voltage spikes when interlock signals fired}\label{label_nineteen}

The behaviour of the system was investigated when the interlock signals were fired for only some of the channels on a power supply module. The aim was to investigate if voltage spikes could occur on other channels on the module.
The interlock signals are provided to the 16 channels on a power supply module via two 9 D-sub connectors in the rear panel of the module. Each connector controls 8 channels.

A crate containing 9 high voltage modules was used and -140V was applied to all channels. The voltage and current hardware limits were set respectively to 700V and 4mA. A channel was chosen for monitoring and the interlock of other channels was fired.  The channel for which the interlock was fired was connected directly to one channel of an oscilloscope and used as the trigger, with the trigger level set at -138V. The monitored channel was connected to another channel of the same oscilloscope. 

\begin{figure}[!ht]
\epsfig{file=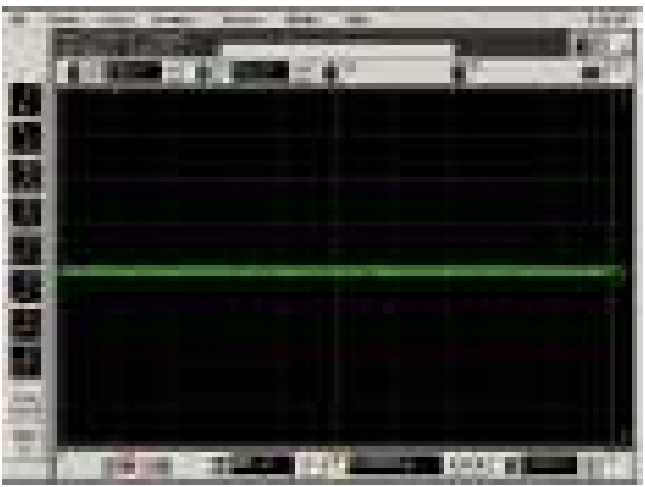,width=0.50\textwidth}
\caption[Oscilloscope trace of a test for voltage spikes when the interlock signals are disabled.] {Oscilloscope trace of a test for voltage spikes when the interlock signals are disabled. Half of the interlock signals of one HV module were fired and another channel monitored. The Voltage scale is 500mV/division. The time scale is 500ns/division.}
\label{picture_seventeen}
\end{figure}

The interlock signals for one half of the power supply module were then fired (eight channels), while a channels on the other eight was monitored. Figure \ref{picture_seventeen} shows that the behaviour of the monitored channel is not affected by firing half of the interlock signals.

\end{itemize}

This test was repeated interrupting the interlock signal for one channel and monitoring the behaviour of its neighbour. Again, this test has shown that firing the interlock signal to one channel does not affect the voltage supplied to its neighbouring channel.

\subsubsection{Speed of voltage cut}\label{label_twenty_two}

Measurements were made to determine the time taken for the voltage to be cut after the interlock signal was fired. During this measurement the channel was connected to the oscilloscope, through a voltage divider, with the following settings:
\begin{itemize}
\item Voltage of -500V was applied.
\item The oscilloscope was in single trigger mode.
\item The trigger level set at -24 V which corresponds to -485 V after compensation for the voltage divider.
\end{itemize}

\begin{figure}[!ht]
\epsfig{file=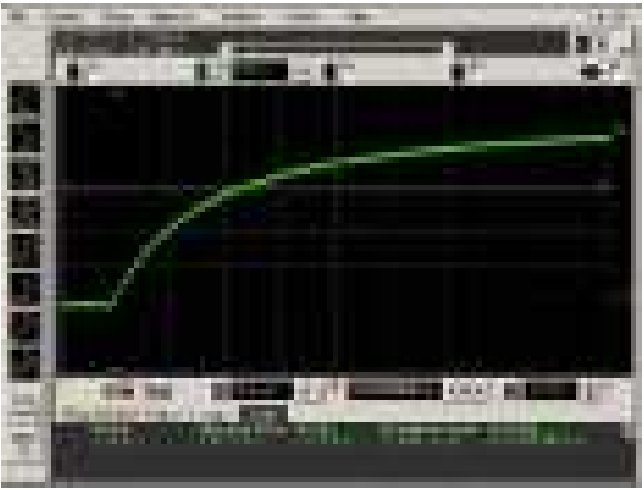,width=0.50\textwidth}
\caption[Oscilloscope trace of the output voltage when the interlock signals were removed] {Oscilloscope trace of the output voltage when the interlock signals were removed. The time scale is 10ms/division.}
\label{picture_eighteen}
\end{figure}

The interlock signals were then fired. Figure \ref{picture_eighteen} shows that it takes approximately 90ms for the module to ramp from -500V to 0V when the HV inhibit is fired. 

This sudden interruption of bias voltage to the sensors could cause a voltage spike at the input of the AC-coupled front-end chips. Accidental power interruptions have occurred during test beam operation from voltages of over 100V and no damage has been observed. However, a test has not been performed from 500V.

%% file: failure_modes/currentlimit.tex
\subsection{Voltage and current limits}

Limits on the maximum voltage and current that can be applied can be set for the power supply module. This section discuses the testing of these limits.

The power supply module can supply up to 500V but its maximum voltage can be restricted through altering a potentiometer. The initial operation of the VELO sensors will be at around 100V, but after radiation damage this voltage is expected to be increased up to a maximum of 500V. Hence, for safety, this potentiometer will be adjusted to limit the applied voltage. 

Similarly, the power supply module can supply up to 4mA. However, these current levels will only be required when the sensors are heavily irradiated. Hence, in early operation the current limit should, for safety, be set considerably lower. The power supply can be set in modes to either `trip' when the maximum current is reached or to limit the supplied current in 'control' mode. 

The maxmum voltage and current are adjusted by the potentiometers, labeled Vmax and Imax, next to the two LEMO connectors on the front panel of the module \cite{HVP}. The potentiometers affect the setting values in the module's microprocessor. The values set by the potentiometers are converted via an ADC-converter and are processed by this microprocessor. The microprocessor limits the maximum value of current or voltage that will be applied. The potentiometer position limits the voltage or current for all channels on the module.

It is possible to measure the hardware current or voltage limit that has been set on the potentiometer through the LEMO connector: The maximal value of Imax (4mA) and Vmax (700V) corresponds to 2.5V measured from the LEMO connector. The linearity of the measured 0-2.5V signal to the current and voltage limits set are discussed below.

In addition to the `hardware' limit on the current a software limit can be applied to each individually channel. This is an additional safety feature. As the current-voltage characteristics of the sensors differ significantly \cite{burn-in} the variability for each channel is also a useful feature.

The power supply module can be operated in two modes: current control or current trip. In the current control mode the maximum current delivered by the system is limited to the value that has been set. In the current trip mode, a maximum safe current is set and the voltage is cut if this value is reached. For the VELO system we recommend using the current trip mode, using this limit as a safety system. The trip value should be set to a value above that which would be reached in normal operation.
The mode of operation is controlled through the software by the `kill enable' bit that can be configured to the values 0 and 1. If the kill enable is set to 0 the mode of operation is current control and if it is 1 the mode is current trip. The choice of mode applies to both 'hardware' and 'software' current limits.
The following tests aim to check that the limits function and the accuracy with which the hardware can be set.

\subsubsection{Voltage `hardware' limit}\label{label_three}

The performance of the voltage limit was tested over the full voltage range of the power supply and was found to function correctly.

\begin{figure}[!ht]
\epsfig{file=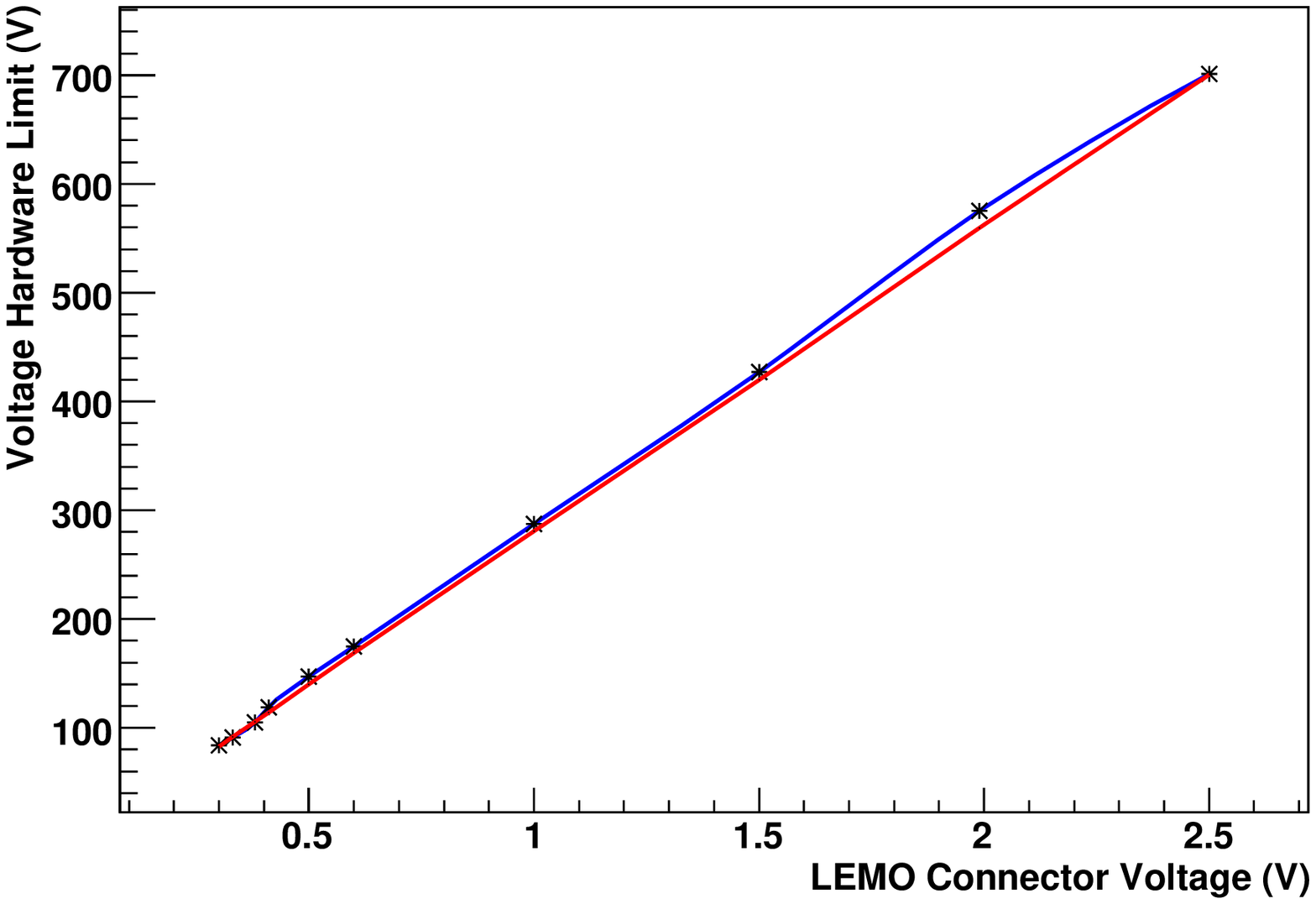,width=0.50\textwidth}
\epsfig{file=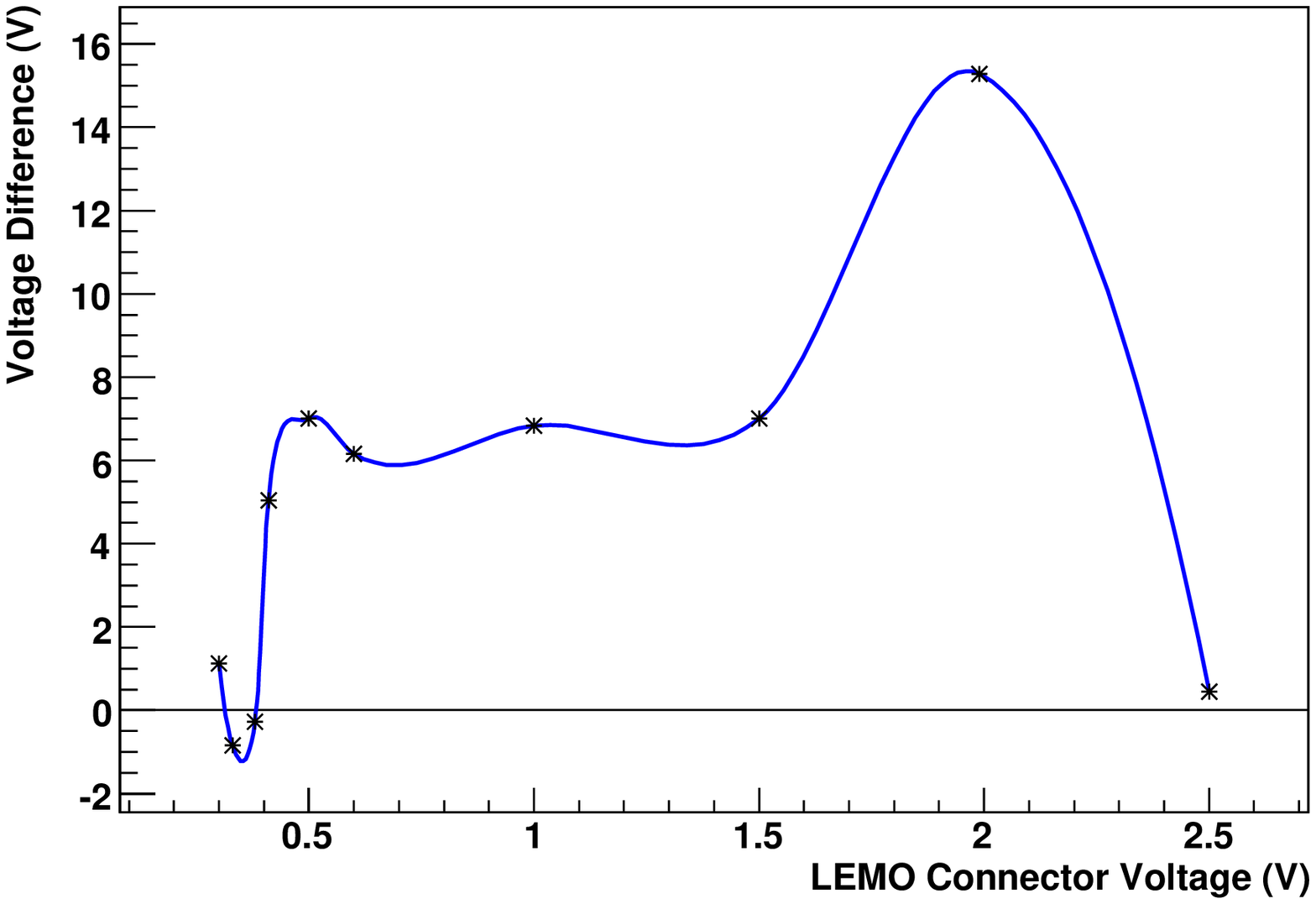,width=0.50\textwidth}
\caption[Characterization of voltage hardware limits.] {A- (left) Hardware voltage limits as a function of voltages measured from the LEMO connector.
B- (right) Discrepancy between the measured hardware voltage limits and linear behaviour for the voltages measured from the LEMO connector.}
\label{picture_seven}
\end{figure}

The linearity between the voltage measured on the LEMO connector (0 to 2.5V) and the voltage limit applied (0 to 700V) was tested. The potentiometer was changed over the full range and thus the voltage measured from the LEMO connector changed accordingly. For each setting the hardware voltage limit was measured. The measured hardware limits as a function of the voltage measured on the LEMO connector are shown in Figure \ref{picture_seven}A as the blue line. The pink line in
Figure \ref{picture_seven}A is given by assuming a linear relationship, where 2.5V corresponds to 700V. The deviation of the measured behaviour from linear is shown in Figure \ref{picture_seven}. The hardware limit has been tested from 84V up to 700V. The discrepancy from linear behaviour should be considered when determining the maximum voltage to set.

\subsubsection{Current `hardware' limit}\label{label_four}

The corresponding tests considered in the previous section for the voltage limit were also applied to the current limit. The current limit was provoked by connecting a resistance of 100k$\Omega$ to the output of the power supply module and ramping the applied voltage. The hardware current limit has been tested from 120$\mu$A up to 4mA.

\begin{figure}[!ht]
\epsfig{file=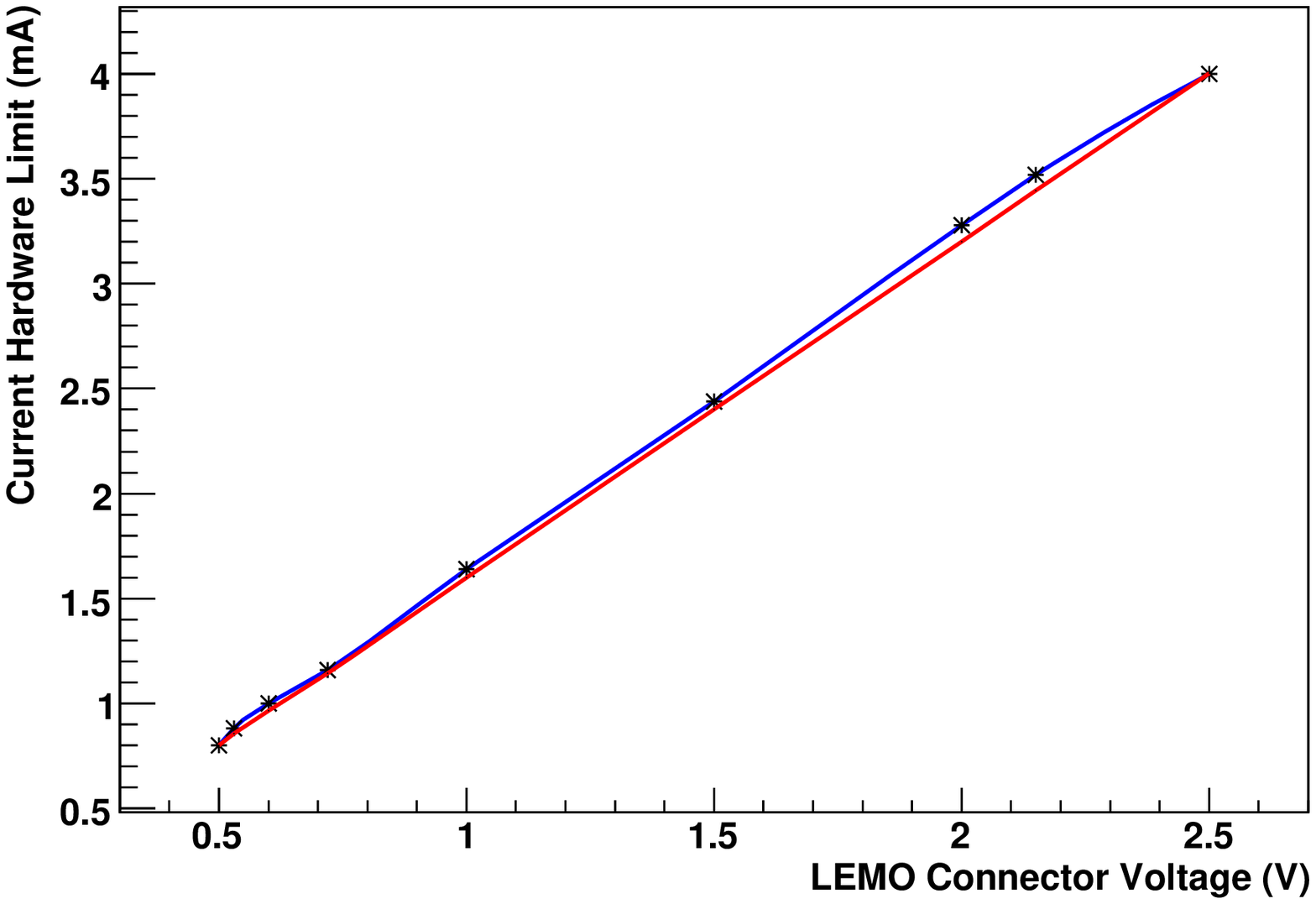,width=0.50\textwidth}
\epsfig{file=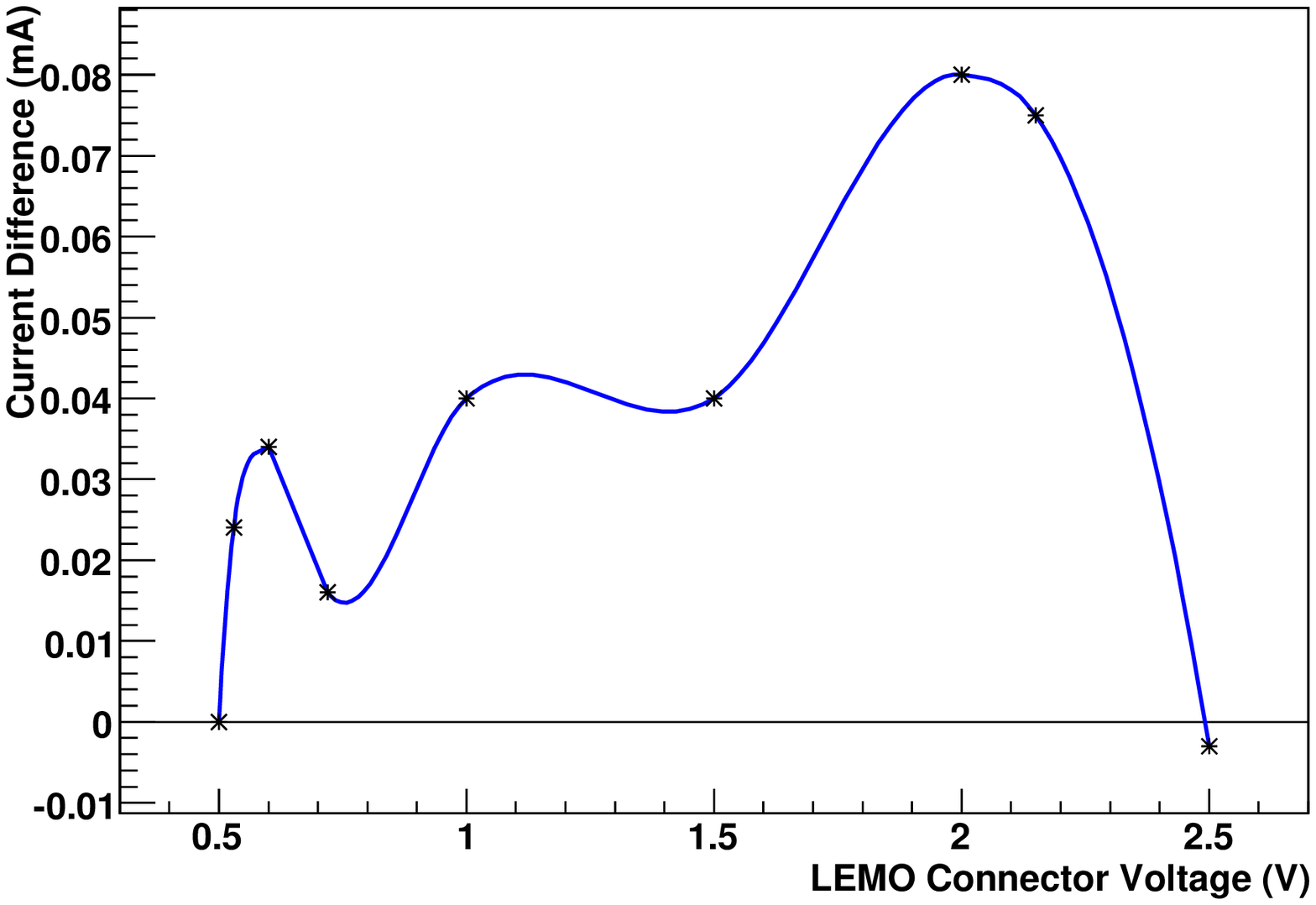,width=0.50\textwidth}
\caption[Characterization of current hardware limits.] {A- (left) Hardware current limits as 
a function of voltages measured from the LEMO connector.
B- (right) Discrepancy between the measured hardware current limits and linear behaviour for the voltages measured from the LEMO connector.}
\label{picture_nine}
\end{figure}

Figure \ref{picture_nine}A shows the measured current limits as a function of them measured voltage on the Imax LEMO connector as the blue line. The pink line in Figure \ref{picture_nine}A is obtained by assuming a linear relationship where 2.5V corresponds to 4mA output on the power supply. Figure \ref{picture_nine}B shows the discrepancy between from the linear behaviour is small.

\subsubsection{Current software limit up to 4mA}\label{label_six}
In addition to the 'hardware' limit that can be set through the potentiometer, software limits on the current can also be set. This test aims to verify that the current software limit of individual channel works. 

The current limit was provoked by applying various voltages to a resistance of 100k$\Omega$. The software limit was tested from 120$\mu$A up to 4mA. The test were performed both in current control mode and in current trip mode. When the software current limit is reached a trip error alarm is displayed in the software, and the current is either limited to the limit value (control mode) or the voltage to this channel turned off (trip mode). After repairing the error, the alarm can be cleared and the channel returned to normal operation. All channels have been tested and for each one the current software limits work correctly.

\subsubsection{Current limit below 120$\mu$A}\label{label_seven}

In the sections above the current limit was only considered above 120$\mu$A. Test were also performed at lower values. Again the current limit was provoked by applying various voltages to a resistance of 100k$\Omega$.
The results show that a ramping speed of 7V/s (i.e. 1\% of full output per second) always provokes a trip for a current software limit below 40$\mu$A even if nothing is connected to the module. If the ramping speed is reduced, e.g. to 1V/s, or the current software limit is set higher this problem is prevented. While it is not expected that the hardware limit would be set below 40$\mu$A, the users need to be aware of this behaviour for setting the software limits.

\subsubsection{Voltage cut in current trip mode}\label{label_twenty_three}

The current trip was designed to shut off the module immediately if activated. The firmware of the modules does not offer the possibility to “ramp down” the channels in the case of trips. This ramp down period would have to be short for safety purposes but could be longer than the current behaviour which is equivalent to that reported in section \ref{label_twenty_two}, where the potential danger of a sudden voltage cut to the modules is considered. The possibility of a future firmware change is not excluded.

%% file: cables_and_patch_panels_testing/cables_and_patch_panels_testing.tex
\section{Cables and patch panels testing}\label{cables_and_patch_panels_testing}
The cables and patch panels of the high voltage system have been tested, including all spare components.

Connectivity tests were performed by the Glasgow team prior to installation on the following components: the counting house cable which links the power supply modules to the patch panel in the counting house; the counting house patch panels; the patch panel close to the detector. A full chain test of these components was then performed up to 500V, using a prototype long distance cable. The leakage current measured was consistent to zero within the resolution of the measurement.
%to ensure there was no leakage current in any of the channels.

The connectivity of the long distance cables that link the counting house and detector patch panels were tested after the in-situ installation of the connectors. One of the connectors was subsequently damaged during the installation of the low voltage system and this connector was replaced and retested.

The detector cables that link the detector patch panels and the HV connectors on the VELO repeater boards were assembled in the two required lengths and tested by LEMO S.A.\footnote{LEMO S.A., CH-1024 Ecublens, Switzerland }.

The connectivity and leakage current at 500V was tested on all channels of the full assembled HV system in the pit, including the spare detector and long distance cables. The description of these tests is given in \cite{HVProcedure}.

%% file: low_voltage_test/low_voltage_test.tex
\section{Low voltage behaviour and current fluctuations}

\subsection{Low voltage behaviour}\label{low_voltage_test}
The high voltage system of the VELO will typically be operated at voltages in excess of 100V. However, during tests it can be useful to apply low voltages to the system, for example, when testing connectivity. During these tests of the system it was observed that the behaviour of the power supply modules at low voltages (below 10V) was not as expected. This behaviour is reported on in this section.

\begin{figure}[!ht]
\epsfig{file=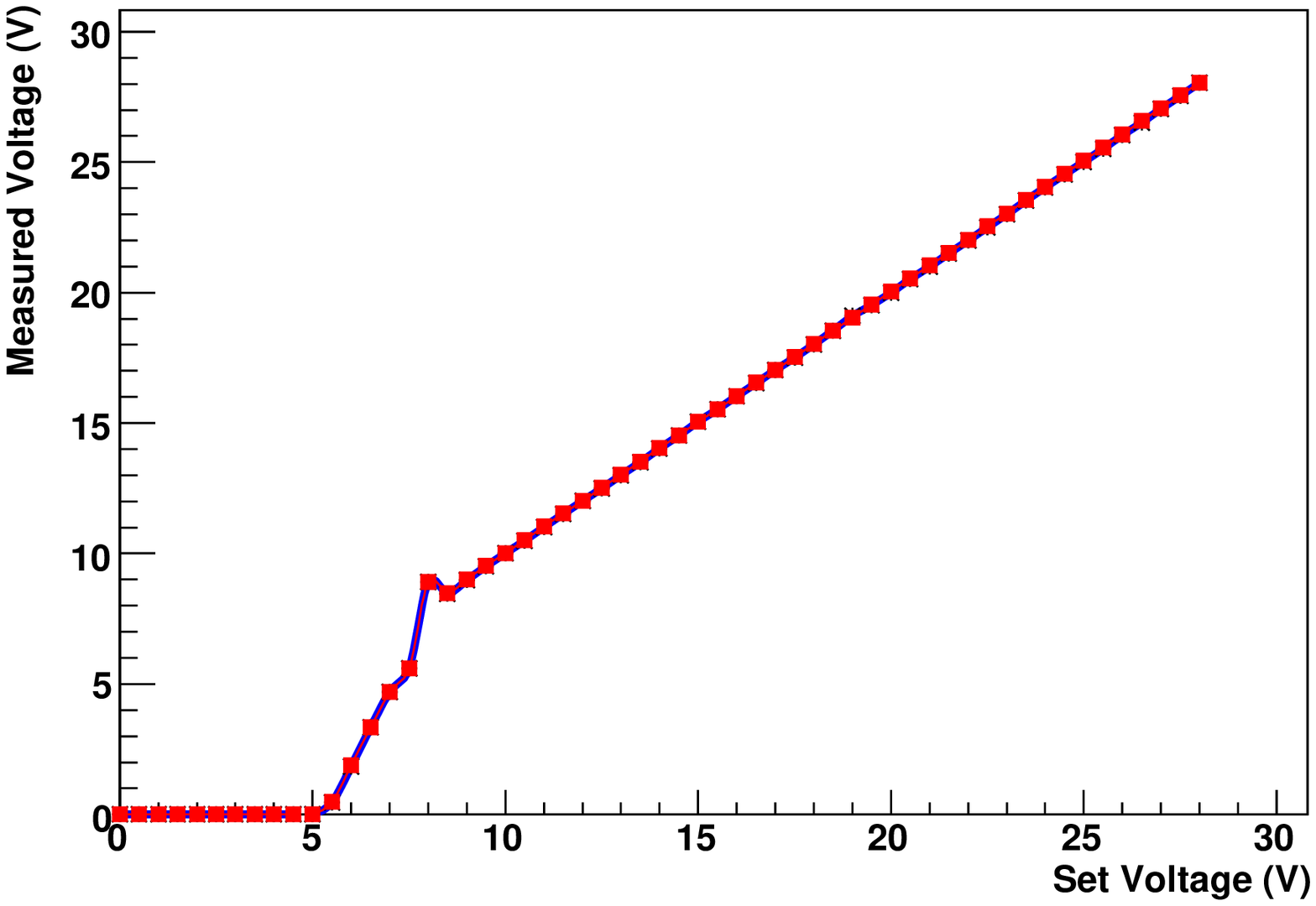,width=0.50\textwidth}
\epsfig{file=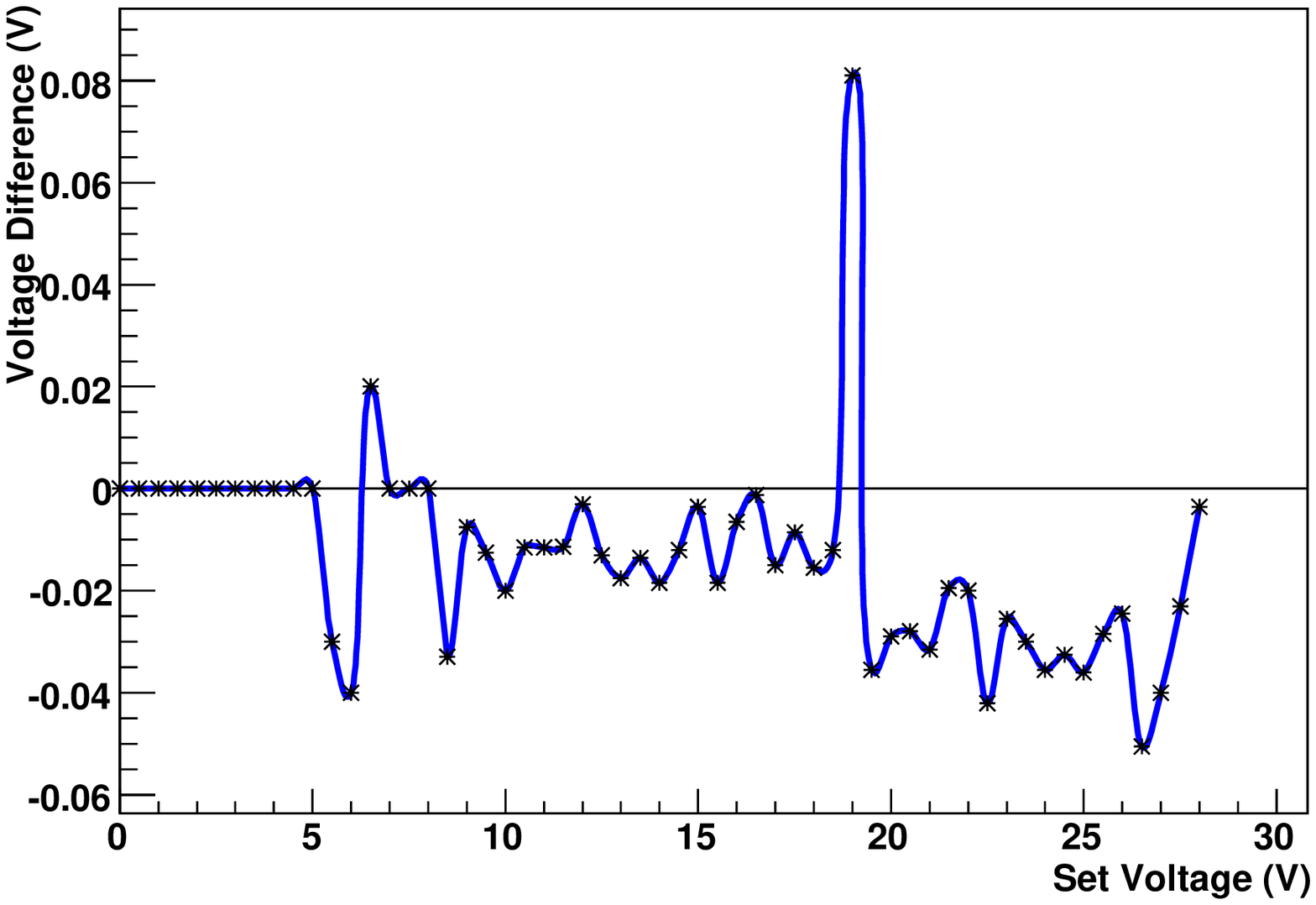,width=0.50\textwidth}
\caption[Comparison of the requested and measured voltages at low voltages] {Comparison of the requested and measured voltages at low voltages. A- (left)  Voltage measured as a function of the set voltage. The blue points, largely obscured, are for PVSS software measurements and the pink points are for DVM measurements. B- (right) Difference between the PVSS and DVM measurements as a function of the requested voltage. These results are for channel 5 of the module with serial number: 474189 07/2006.}
\label{picture_nineteen}
\end{figure}

The requested value of the high voltage was compared with the voltage obtained as read back from the system and displayed in the software and cross-checked with a DVM. Figure \ref{picture_nineteen}A shows the result obtained when setting voltages from 0V to 28V via the PVSS software in steps of 0.5V. The PVSS measurements (blue points) and the DVM measurements (pink points) are all averages over several measurements. Both measurements are consistent (indeed the blue points are largely obscured by the pink points on this scale) and their differences
are presented in Figure \ref{picture_nineteen}B.

It is clear from Figure \ref{picture_nineteen} that the behaviour above 9V is as expected, the measured voltage in software and through the DVM both agree we the voltage that has been requested. This confirms the normal operating behaviour that was reported in section \ref{normal_operation}. 

\begin{figure}[!ht]
\epsfig{file=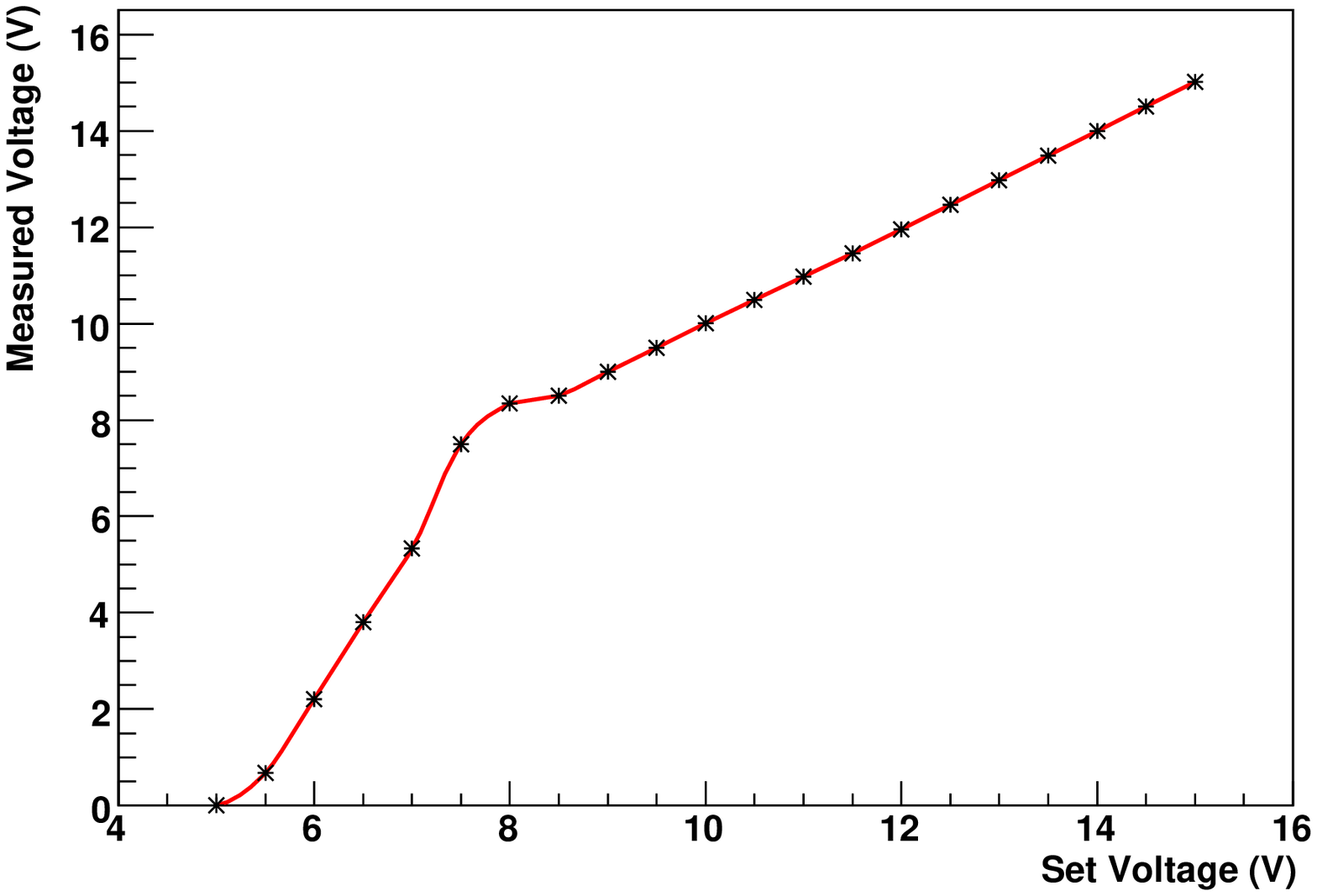,width=0.50\textwidth}
\epsfig{file=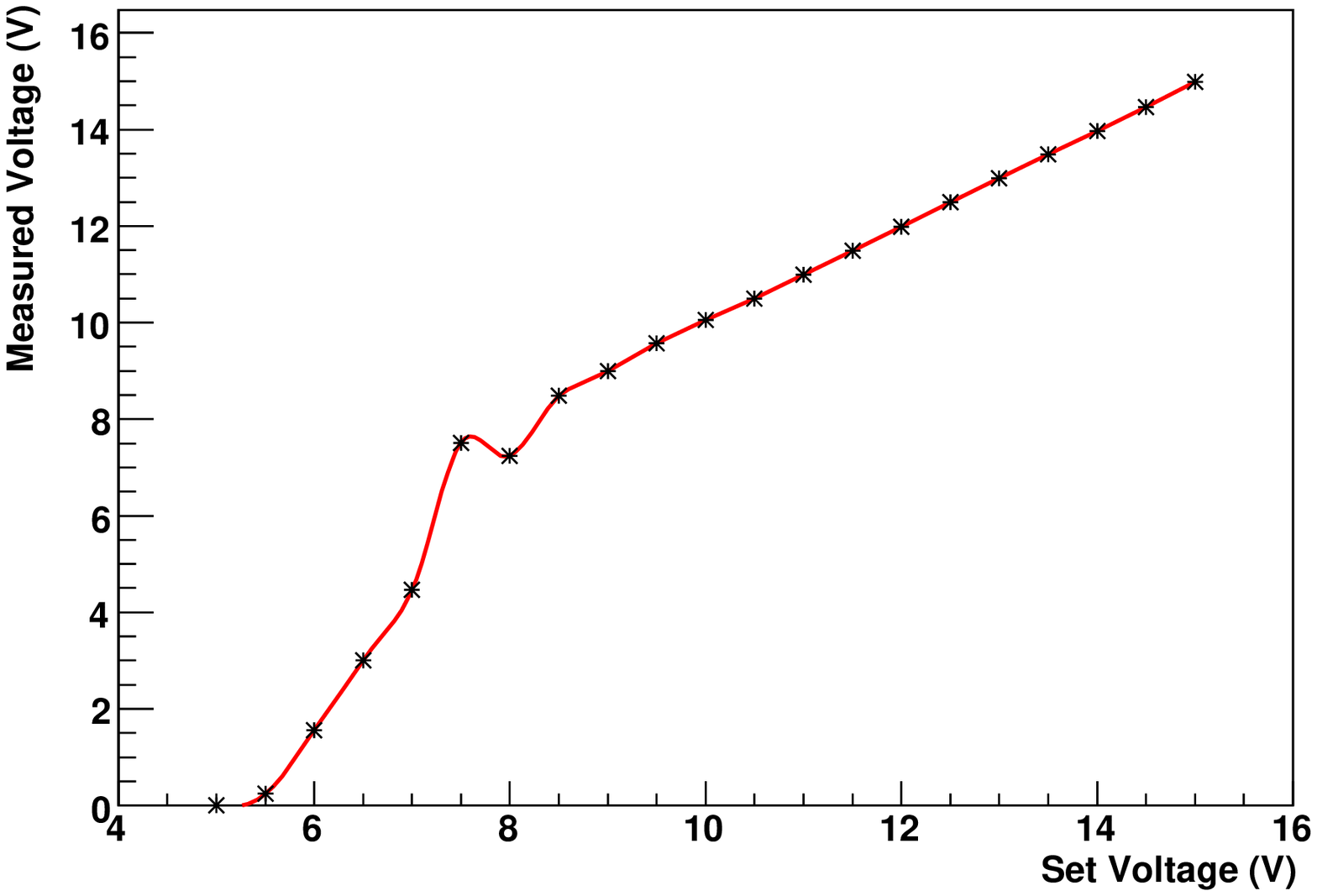,width=0.50\textwidth}
\epsfig{file=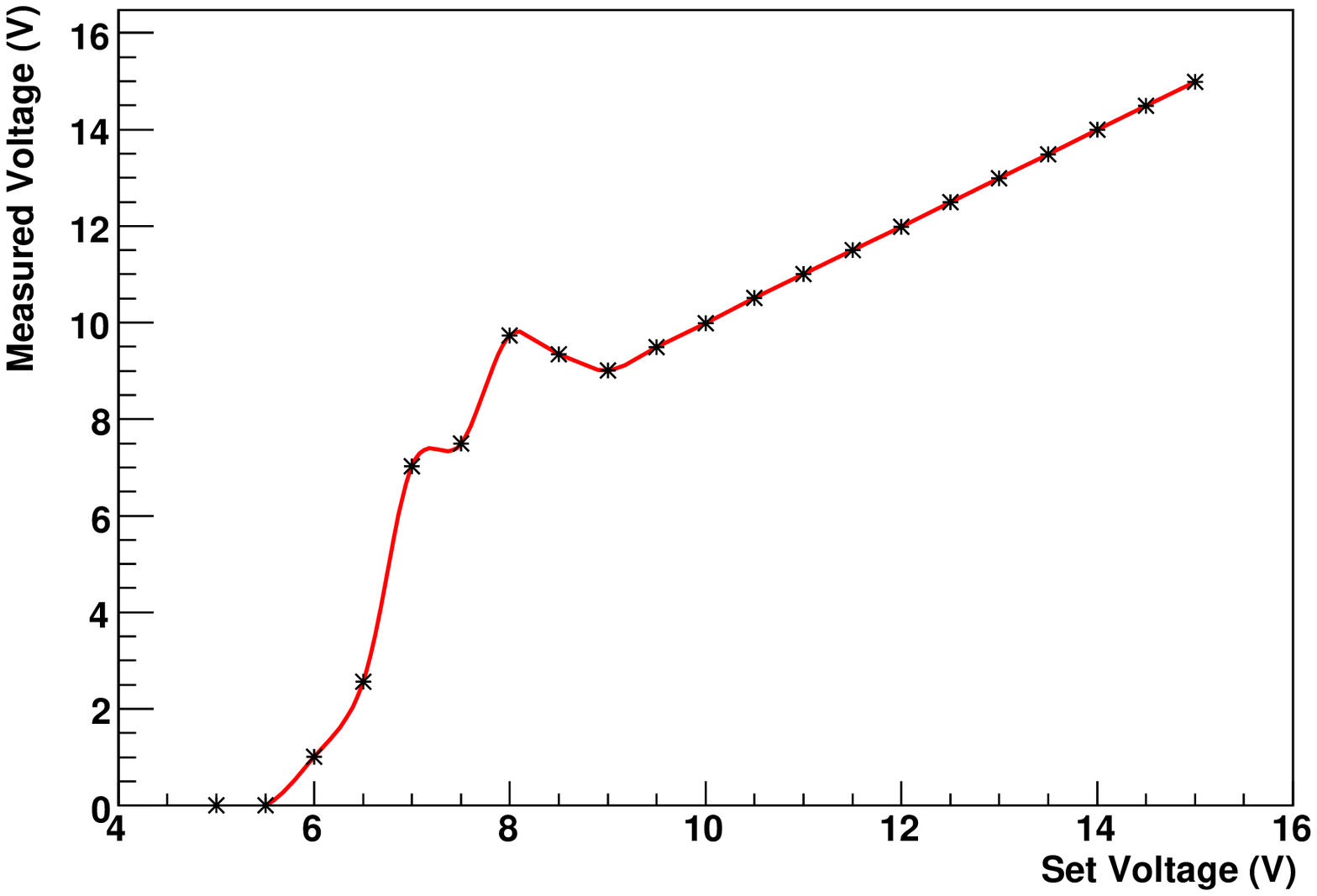,width=0.50\textwidth}
\epsfig{file=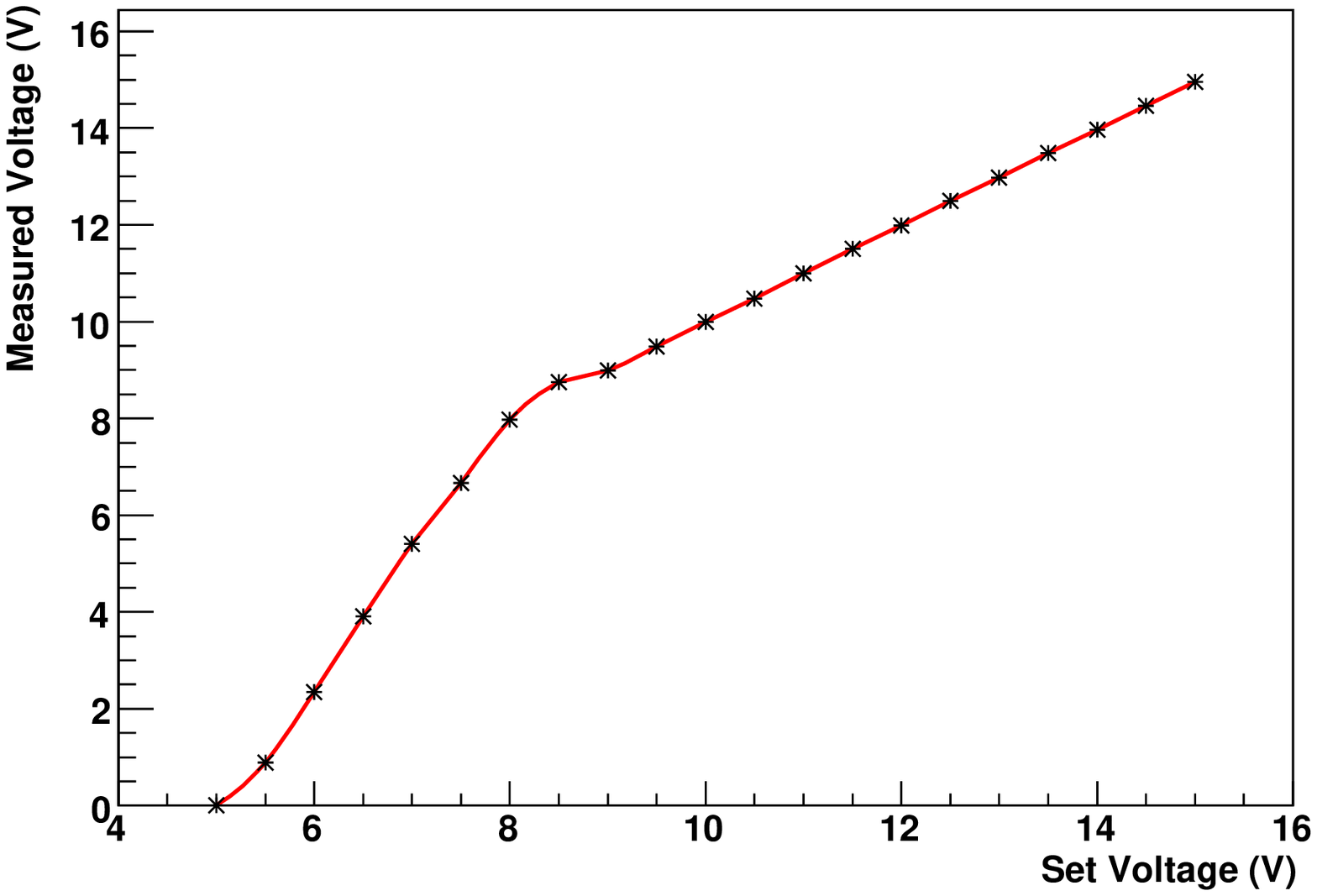,width=0.50\textwidth}
\caption[Measured voltage as a function of set voltage at low voltages for several channels] {Measured voltage as a 
function of set voltage for low voltages. 
A- (top left) channel 1 of the module with SN: 474195 07/2006
B- (top right) channel 10 of the module with SN: 474193 07/2006
C- (bottom left) channel 15 of the module with SN: 474191 07/2006
D- (bottom right) channel 2 of the module with SN: 474191 07/2006}
\label{picture_ninefive}
\end{figure}

However, below 9V the voltage applied does not correspond to the voltage that has been set. Further measurements have been taken from channels of several high voltage modules to understand its behaviour in this non linear part of the graph.
Figure \ref{picture_ninefive} reports the behaviour of four channels. These results are representative of the larger sample of measurements that were made. 
The following conclusions are drawn:
\begin{itemize}
\item The high voltage module operates correctly for voltage settings above 9V.
\item Setting a voltage below 5V, results in no voltage being applied.
\item For requested voltages between 5V and 6.5V the voltage applied is less than the voltage set.
\item For requested voltages between 6.5V and 9V the voltage applied is typically higher than that which has been set. The highest voltage set is approximately 10V.
\end{itemize}

Note also that the behaviour of all channels on a single module is not the same, see \ref{picture_ninefive}C and D. It was not clear that the variation in behaviour between channels on different modules (e.g. \ref{picture_ninefive}A, B, C) is any larger than between channels on the same module.

The difference between the values set and the voltage measured in this low voltage region are thought to be due to the noise of the ADC and of the voltage dividers in the power supply module in this region. In discussion with the power supply company, the modules were claimed to typically be calibrated above 1\% of the maximum voltage, i.e. above 7V. However, the company has seen problems with the calibration and suggest that the true calibration may be only achieved above 1.5\% (11V) to 2\% (14V). From the tests reported here, we advise only using the system above 10V.

\subsection{Current reading fluctuation}\label{current_fluctuation}

The accuracy of the current reading of the ISEG module has been tested. The ISEG manual \cite{HVP} states that the current measurement resolution is 80nA
and the accuracy of the measurement is given as 
$\emph{0.005 \% I + 0.01 \% Imax}$, where $\emph{I}$ is the supplied current and $\emph{Imax}$ is 4mA (the maximum current the power supply can produce), so for low currents the precision is 0.4$\mu$A r.m.s.
The fluctuations in measured current values with stable loads have been observed. This was tested for all channels of the high voltage system without any load applied at 100V. The test was repeated for a few number of channels per module with a 100M$\Omega$ resistance and operated at 100V. The results of both tests are in agreement.

Fluctuations of 4$\mu$A peak-to-peak amplitude were observed on power supply module 474187 07/2006. 
This is considerably worse than observed on the other modules, which all showed similar behaviour and similar to that expected from the manual. 
The peak-to-peak amplitude of the fluctuation observed is always less than 1$\mu$A, for all the modules used in the VELO system. The fluctuations are caused by a measurement error, they are not actual variations in the supplied current this has been checked with an oscilloscope and a DVM.
The voltage was set and measured using the Iseg standalone control software IsegHVControl version 1.57. 
In addition, a comparison has been made by using PVSS, using the Iseg OPC server 5.00.006.

As typical currents drawn by the unirradiated modules are less than 5$\mu$A at 250V, a fluctuation of 1$\mu$A is of relevance for early operation: the currents after irradiation will be considerably higher. Indeed, the current reading provides useful information on the level of irradiation. The level of the observed fluctuations are under discussion with Iseg. 
The accuracy of the measurement of IV curves on the VELO modules has been improved by implementing an averaging procedure in the PVSS FSM. The current values are recorded ten times and the average value used in the IV plots produced.

%\clearpage

%% file: summary/summary.tex
\section{Individual test summary}\label{summary}
%%Successful example to color rows in a table njaka 20-03-08 
%
%\begin{tabular}{|c|c|}
 %\multicolumn{1}{|>{\columncolor[named]{VioletRed}}r|}{0}  & \multicolumn{1}{|>{\columncolor[named]{SeaGreen}}r|} {1} \\
 %\hline
 %\end{tabular}

This section details to which channel of each module each test described in this document was applied. These are specified in tables 1 to 9.
% Check how to make reference to table 1 to 6
Some basic functionality tests were applied to every module, others were applied to a sample of channels.
Modules 1, 2, 3, 4, 5, 6 are currently used in the experiment. Modules 7,8,9 are spares. The key to the colour coding of these tables 1 to 9 is as follows:
\\   \\
\begin{tabular}{l l}
%\ref{label_nineteen}A & \multicolumn {1}{l}{} \\
%\ref{label_nineteen}B & \multicolumn {1}{l}{} \\
%\ref{label_nineteen}C & \multicolumn {1}{l}{Interlock on this module was fired} \\
\multicolumn{1}{|>{\columncolor[named]{Green}}r}{} & Tested, no problem found \\
\multicolumn{1}{|>{\columncolor[named]{Gray}}r}{} & Measurement made, results in the corresponding section \\
\multicolumn{1}{|>{\columncolor[named]{Orange}}r}{} & The interlock was fired on this channel as part of this test\\
\multicolumn{1}{|>{\columncolor[named]{Cerulean}}r}{} & The interlock was fired on this channel and in addition\\ 
 & the oscilloscope was triggered on this channel as part of this test \\
\end{tabular}
\\ \\
The test described in section \ref{label_nineteen} has three components which have been labeled A,B,C in the table below. A refers to the test when half of the interlocks on this module were fired. B to the test when the interlock of one of the neighbouring channel was fired. C indicates the test when all the interlocks on the module were fired.
\clearpage
\input{summary/part_1.tex}
\input{summary/part_2.tex}
\input{summary/part_3.tex}
\input{summary/part_4.tex}
\input{summary/part_5.tex}
\input{summary/part_6.tex}
\input{summary/part_7.tex}
\input{summary/part_8.tex}
\input{summary/part_9.tex}

%% file: summary/part_1.tex
%%% Full Module 01
\begin{table}[h]
\begin{tabular}{|c|c|c|c|c|c|c|c|c|c|c|c|c|c|c|c|c|}
\hline
HV module & \multicolumn {16}{|c|}{EHQ F607-nF} \\
\hline
Serial number & \multicolumn {16}{|c|}{474187 07/2006, Module 1} \\
\hline
Point 8 number & \multicolumn {16}{|c|}{4TVGLAHVPSM01} \\
\hline
Individual test &  \multicolumn {16}{|c|}{Channel number} \\
\hline
%3.1
\ref{label_one} & \multicolumn{1}{|>{\columncolor[named]{Green}}r|}{0} & \multicolumn{1}{|>{\columncolor[named]{Green}}r|}{1} &
 \multicolumn{1}{|>{\columncolor[named]{Green}}r|}{2} & \multicolumn{1}{|>{\columncolor[named]{Green}}r|}{3} &
 \multicolumn{1}{|>{\columncolor[named]{Green}}r|}{4} & \multicolumn{1}{|>{\columncolor[named]{Green}}r|}{5} &
 \multicolumn{1}{|>{\columncolor[named]{Green}}r|}{6} & \multicolumn{1}{|>{\columncolor[named]{Green}}r|}{7} &
 \multicolumn{1}{|>{\columncolor[named]{Green}}r|}{8} & \multicolumn{1}{|>{\columncolor[named]{Green}}r|}{9} &
 \multicolumn{1}{|>{\columncolor[named]{Green}}r|}{10} & \multicolumn{1}{|>{\columncolor[named]{Green}}r|}{11} &
 \multicolumn{1}{|>{\columncolor[named]{Green}}r|}{12} & \multicolumn{1}{|>{\columncolor[named]{Green}}r|}{13} &
 \multicolumn{1}{|>{\columncolor[named]{Green}}r|}{14} & \multicolumn{1}{|>{\columncolor[named]{Green}}r|}{15} \\
\hline

\end{tabular}
\\
\caption[Summary of tests performed on Module 1]{Summary of tests performed on Module 1 (see text).}
\end{table}
\clearpage

%% file: summary/part_2.tex
%%%%% Full Module 02
\begin{table}[h]
\begin{tabular}{|c|c|c|c|c|c|c|c|c|c|c|c|c|c|c|c|c|}
\hline
HV module & \multicolumn {16}{|c|}{EHQ F607-nF} \\
\hline
Serial number & \multicolumn {16}{|c|}{474189 07/2006, Module 2} \\
\hline
Point 8 number & \multicolumn {16}{|c|}{4TVGLAHVPSM02} \\
\hline
Individual test &  \multicolumn {16}{|c|}{Channel number} \\
\hline
%3.1
\ref{label_one} & \multicolumn{1}{|>{\columncolor[named]{Green}}r|}{0} & \multicolumn{1}{|>{\columncolor[named]{Green}}r|}{1} &
 \multicolumn{1}{|>{\columncolor[named]{Green}}r|}{2} & \multicolumn{1}{|>{\columncolor[named]{Green}}r|}{3} &
 \multicolumn{1}{|>{\columncolor[named]{Green}}r|}{4} & \multicolumn{1}{|>{\columncolor[named]{Green}}r|}{5} &
 \multicolumn{1}{|>{\columncolor[named]{Green}}r|}{6} & \multicolumn{1}{|>{\columncolor[named]{Green}}r|}{7} &
 \multicolumn{1}{|>{\columncolor[named]{Green}}r|}{8} & \multicolumn{1}{|>{\columncolor[named]{Green}}r|}{9} &
 \multicolumn{1}{|>{\columncolor[named]{Green}}r|}{10} & \multicolumn{1}{|>{\columncolor[named]{Green}}r|}{11} &
 \multicolumn{1}{|>{\columncolor[named]{Green}}r|}{12} & \multicolumn{1}{|>{\columncolor[named]{Green}}r|}{13} &
 \multicolumn{1}{|>{\columncolor[named]{Green}}r|}{14} & \multicolumn{1}{|>{\columncolor[named]{Green}}r|}{15} \\
\hline

\end{tabular}
\\
\caption[Summary of tests performed on Module 2]{Summary of tests performed on Module 2 (see text).}
\end{table}
\clearpage

%% file: summary/part_3.tex
%%%%%% Full Module 03
\begin{table}[h]
\begin{tabular}{|c|c|c|c|c|c|c|c|c|c|c|c|c|c|c|c|c|}
\hline
HV module & \multicolumn {16}{|c|}{EHQ F607-nF} \\
\hline
Serial number & \multicolumn {16}{|c|}{474185 07/2006, Module 3} \\
\hline
Point 8 number & \multicolumn {16}{|c|}{4TVGLAHVPSM03} \\
\hline
Individual test & \multicolumn {16}{|c|}{Channel number} \\
\hline
%3.1
\ref{label_one} & \multicolumn{1}{|>{\columncolor[named]{Green}}r|}{0} & \multicolumn{1}{|>{\columncolor[named]{Green}}r|}{1} &
 \multicolumn{1}{|>{\columncolor[named]{Green}}r|}{2} & \multicolumn{1}{|>{\columncolor[named]{Green}}r|}{3} &
 \multicolumn{1}{|>{\columncolor[named]{Green}}r|}{4} & \multicolumn{1}{|>{\columncolor[named]{Green}}r|}{5} &
 \multicolumn{1}{|>{\columncolor[named]{Green}}r|}{6} & \multicolumn{1}{|>{\columncolor[named]{Green}}r|}{7} &
 \multicolumn{1}{|>{\columncolor[named]{Green}}r|}{8} & \multicolumn{1}{|>{\columncolor[named]{Green}}r|}{9} &
 \multicolumn{1}{|>{\columncolor[named]{Green}}r|}{10} & \multicolumn{1}{|>{\columncolor[named]{Green}}r|}{11} &
 \multicolumn{1}{|>{\columncolor[named]{Green}}r|}{12} & \multicolumn{1}{|>{\columncolor[named]{Green}}r|}{13} &
 \multicolumn{1}{|>{\columncolor[named]{Green}}r|}{14} & \multicolumn{1}{|>{\columncolor[named]{Green}}r|}{15} \\
\hline

\end{tabular}
\\
\caption[Summary of tests performed on Module 3]{Summary of tests performed on Module 3 (see text).}
\end{table}
\clearpage

%% file: summary/part_4.tex
%%%% Full Module 04
\begin{table}[h]
\begin{tabular}{|c|c|c|c|c|c|c|c|c|c|c|c|c|c|c|c|c|}
\hline
HV module & \multicolumn {16}{|c|}{EHQ F607-nF} \\
\hline
Serial number & \multicolumn {16}{|c|}{474191 07/2006, Module 4} \\
\hline
Point 8 number & \multicolumn {16}{|c|}{4TVGLAHVPSM04} \\
\hline
Individual test & \multicolumn {16}{|c|}{Channel number} \\
\hline
%3.1
\ref{label_one} & \multicolumn{1}{|>{\columncolor[named]{Green}}r|}{0} & \multicolumn{1}{|>{\columncolor[named]{Green}}r|}{1} &
 \multicolumn{1}{|>{\columncolor[named]{Green}}r|}{2} & \multicolumn{1}{|>{\columncolor[named]{Green}}r|}{3} &
 \multicolumn{1}{|>{\columncolor[named]{Green}}r|}{4} & \multicolumn{1}{|>{\columncolor[named]{Green}}r|}{5} &
 \multicolumn{1}{|>{\columncolor[named]{Green}}r|}{6} & \multicolumn{1}{|>{\columncolor[named]{Green}}r|}{7} &
 \multicolumn{1}{|>{\columncolor[named]{Green}}r|}{8} & \multicolumn{1}{|>{\columncolor[named]{Green}}r|}{9} &
 \multicolumn{1}{|>{\columncolor[named]{Green}}r|}{10} & \multicolumn{1}{|>{\columncolor[named]{Green}}r|}{11} &
 \multicolumn{1}{|>{\columncolor[named]{Green}}r|}{12} & \multicolumn{1}{|>{\columncolor[named]{Green}}r|}{13} &
 \multicolumn{1}{|>{\columncolor[named]{Green}}r|}{14} & \multicolumn{1}{|>{\columncolor[named]{Green}}r|}{15} \\
\hline

\end{tabular}
\\
\caption[Summary of tests performed on Module 4]{Summary of tests performed on Module 4 (see text).}
\end{table}
\clearpage

%% file: summary/part_5.tex
%%%% End Module 05
\begin{table}[h]
\begin{tabular}{|c|c|c|c|c|c|c|c|c|c|c|c|c|c|c|c|c|}
\hline
HV module & \multicolumn {16}{|c|}{EHQ F607-nF} \\
\hline
Serial number & \multicolumn {16}{|c|}{474193 07/2006, Module 5} \\
\hline
Point 8 number & \multicolumn {16}{|c|}{4TVGLAHVPSM05} \\
\hline
Individual test & \multicolumn {16}{|c|}{Channel number} \\
\hline
%3.1
\ref{label_one} & \multicolumn{1}{|>{\columncolor[named]{Green}}r|}{0} & \multicolumn{1}{|>{\columncolor[named]{Green}}r|}{1} &
 \multicolumn{1}{|>{\columncolor[named]{Green}}r|}{2} & \multicolumn{1}{|>{\columncolor[named]{Green}}r|}{3} &
 \multicolumn{1}{|>{\columncolor[named]{Green}}r|}{4} & \multicolumn{1}{|>{\columncolor[named]{Green}}r|}{5} &
 \multicolumn{1}{|>{\columncolor[named]{Green}}r|}{6} & \multicolumn{1}{|>{\columncolor[named]{Green}}r|}{7} &
 \multicolumn{1}{|>{\columncolor[named]{Green}}r|}{8} & \multicolumn{1}{|>{\columncolor[named]{Green}}r|}{9} &
 \multicolumn{1}{|>{\columncolor[named]{Green}}r|}{10} & \multicolumn{1}{|>{\columncolor[named]{Green}}r|}{11} &
 \multicolumn{1}{|>{\columncolor[named]{Green}}r|}{12} & \multicolumn{1}{|>{\columncolor[named]{Green}}r|}{13} &
 \multicolumn{1}{|>{\columncolor[named]{Green}}r|}{14} & \multicolumn{1}{|>{\columncolor[named]{Green}}r|}{15} \\
\hline

\end{tabular}
\\
\caption[Summary of tests performed on Module 5]{Summary of tests performed on Module 5 (see text).}
\end{table}
\clearpage

%% file: summary/part_6.tex
%%%% Full Module 06
\begin{table}[h]
\begin{tabular}{|c|c|c|c|c|c|c|c|c|c|c|c|c|c|c|c|c|}
\hline
HV module & \multicolumn {16}{|c|}{EHQ F607-nF} \\
\hline
Serial number & \multicolumn {16}{|c|}{474195 07/2006, Module 6} \\
\hline
Point 8 number & \multicolumn {16}{|c|}{4TVGLAHVPSMD06} \\
\hline
Individual test &  \multicolumn {16}{|c|}{Channel number} \\
\hline
%3.1
\ref{label_one} & \multicolumn{1}{|>{\columncolor[named]{Green}}r|}{0} & \multicolumn{1}{|>{\columncolor[named]{Green}}r|}{1} &
 \multicolumn{1}{|>{\columncolor[named]{Green}}r|}{2} & \multicolumn{1}{|>{\columncolor[named]{Green}}r|}{3} &
 \multicolumn{1}{|>{\columncolor[named]{Green}}r|}{4} & \multicolumn{1}{|>{\columncolor[named]{Green}}r|}{5} &
 \multicolumn{1}{|>{\columncolor[named]{Green}}r|}{6} & \multicolumn{1}{|>{\columncolor[named]{Green}}r|}{7} &
 \multicolumn{1}{|>{\columncolor[named]{Green}}r|}{8} & \multicolumn{1}{|>{\columncolor[named]{Green}}r|}{9} &
 \multicolumn{1}{|>{\columncolor[named]{Green}}r|}{10} & \multicolumn{1}{|>{\columncolor[named]{Green}}r|}{11} &
 \multicolumn{1}{|>{\columncolor[named]{Green}}r|}{12} & \multicolumn{1}{|>{\columncolor[named]{Green}}r|}{13} &
 \multicolumn{1}{|>{\columncolor[named]{Green}}r|}{14} & \multicolumn{1}{|>{\columncolor[named]{Green}}r|}{15} \\
\hline

\end{tabular}
\\
\caption[Summary of tests performed on Module 6]{Summary of tests performed on Module 6 (see text).}
\end{table}
\clearpage

%% file: summary/part_7.tex
%%%%% Full Module 07
\begin{table}[h]
\begin{tabular}{|c|c|c|c|c|c|c|c|c|c|c|c|c|c|c|c|c|}
\hline
HV module & \multicolumn {16}{|c|}{EHQ F607-nF} \\
\hline
Serial number & \multicolumn {16}{|c|}{474197 07/2006, Module 7} \\
\hline
Point 8 number & \multicolumn {16}{|c|}{Spare} \\
\hline
Individual test & \multicolumn {16}{|c|}{Channel number} \\
\hline
%3.1
\ref{label_one} & \multicolumn{1}{|>{\columncolor{green}}r|}{0}  & \multicolumn{1}{|>{\columncolor{green}}r|} {1} & 
 \multicolumn{1}{|>{\columncolor{green}}r|} {2} & \multicolumn{1}{|>{\columncolor{green}}r|} {3} & 
 \multicolumn{1}{|>{\columncolor{green}}r|} {4} & \multicolumn{1}{|>{\columncolor{green}}r|} {5} &
 \multicolumn{1}{|>{\columncolor{green}}r|} {6} & \multicolumn{1}{|>{\columncolor{green}}r|} {7} & 
 \multicolumn{1}{|>{\columncolor{green}}r|} {8} & \multicolumn{1}{|>{\columncolor{green}}r|} {9} &
 \multicolumn{1}{|>{\columncolor{green}}r|} {10} & \multicolumn{1}{|>{\columncolor{green}}r|} {11} &
 \multicolumn{1}{|>{\columncolor{green}}r|} {12} & \multicolumn{1}{|>{\columncolor{green}}r|} {13} & 
 \multicolumn{1}{|>{\columncolor{green}}r|} {14} & \multicolumn{1}{|>{\columncolor{green}}r|} {15} \\
\hline
%3.2
\ref{label_two} & \multicolumn{1}{|>{\columncolor{green}}r|}{0}  & \multicolumn{1}{|>{\columncolor{green}}r|} {1} &
 \multicolumn{1}{|>{\columncolor{green}}r|} {2} & \multicolumn{1}{|>{\columncolor{green}}r|} {3} &
 \multicolumn{1}{|>{\columncolor{green}}r|} {4} & \multicolumn{1}{|>{\columncolor{green}}r|} {5} &
 \multicolumn{1}{|>{\columncolor{green}}r|} {6} & \multicolumn{1}{|>{\columncolor{green}}r|} {7} &
 \multicolumn{1}{|>{\columncolor{green}}r|} {8} & \multicolumn{1}{|>{\columncolor{green}}r|} {9} &
 \multicolumn{1}{|>{\columncolor{green}}r|} {10} & \multicolumn{1}{|>{\columncolor{green}}r|} {11} &
 \multicolumn{1}{|>{\columncolor{green}}r|} {12} & \multicolumn{1}{|>{\columncolor{green}}r|} {13} &
 \multicolumn{1}{|>{\columncolor{green}}r|} {14} & \multicolumn{1}{|>{\columncolor{green}}r|} {15} \\
\hline

\end{tabular}
\\
\caption[Summary of tests performed on Module 7]{Summary of tests performed on Module 7 (see text).}
\end{table}
\clearpage

%% file: summary/part_8.tex
%%%%% Start Module 08
\begin{table}[h]
\begin{tabular}{|c|c|c|c|c|c|c|c|c|c|c|c|c|c|c|c|c|}
\hline
HV module & \multicolumn {16}{|c|}{EHQ F607-nF} \\
\hline
Serial number & \multicolumn {16}{|c|}{474199 07/2006, Module 8} \\
\hline
Point 8 number & \multicolumn {16}{|c|}{Spare} \\
\hline
Individual test &  \multicolumn {16}{|c|}{Channel number} \\
\hline
%3.1
\ref{label_one} & \multicolumn{1}{|>{\columncolor[named]{Green}}r|}{0} & \multicolumn{1}{|>{\columncolor[named]{Green}}r|}{1} &
 \multicolumn{1}{|>{\columncolor[named]{Green}}r|}{2} & \multicolumn{1}{|>{\columncolor[named]{Green}}r|}{3} &
 \multicolumn{1}{|>{\columncolor[named]{Green}}r|}{4} & \multicolumn{1}{|>{\columncolor[named]{Green}}r|}{5} &
 \multicolumn{1}{|>{\columncolor[named]{Green}}r|}{6} & \multicolumn{1}{|>{\columncolor[named]{Green}}r|}{7} &
 \multicolumn{1}{|>{\columncolor[named]{Green}}r|}{8} & \multicolumn{1}{|>{\columncolor[named]{Green}}r|}{9} &
 \multicolumn{1}{|>{\columncolor[named]{Green}}r|}{10} & \multicolumn{1}{|>{\columncolor[named]{Green}}r|}{11} &
 \multicolumn{1}{|>{\columncolor[named]{Green}}r|}{12} & \multicolumn{1}{|>{\columncolor[named]{Green}}r|}{13} &
 \multicolumn{1}{|>{\columncolor[named]{Green}}r|}{14} & \multicolumn{1}{|>{\columncolor[named]{Green}}r|}{15} \\
\hline

\end{tabular}
\\
\caption[Summary of tests performed on Module 8]{Summary of tests performed on Module 8 (see text).}
\end{table}
\clearpage

%% file: summary/part_9.tex
%%%% Full Module 09 
\begin{table}[h]
\begin{tabular}{|c|c|c|c|c|c|c|c|c|c|c|c|c|c|c|c|c|}
\hline
HV module & \multicolumn {16}{|c|}{EHQ F607-nF} \\
\hline
Serial number & \multicolumn {16}{|c|}{474201 07/2006, Module 9} \\
\hline
Point 8 number & \multicolumn {16}{|c|}{Spare} \\
\hline
Individual test &  \multicolumn {16}{|c|}{Channel number} \\
\hline
%3.1
\ref{label_one} & \multicolumn{1}{|>{\columncolor[named]{Green}}r|}{0} & \multicolumn{1}{|>{\columncolor[named]{Green}}r|}{1} &
 \multicolumn{1}{|>{\columncolor[named]{Green}}r|}{2} & \multicolumn{1}{|>{\columncolor[named]{Green}}r|}{3} &
 \multicolumn{1}{|>{\columncolor[named]{Green}}r|}{4} & \multicolumn{1}{|>{\columncolor[named]{Green}}r|}{5} &
 \multicolumn{1}{|>{\columncolor[named]{Green}}r|}{6} & \multicolumn{1}{|>{\columncolor[named]{Green}}r|}{7} &
 \multicolumn{1}{|>{\columncolor[named]{Green}}r|}{8} & \multicolumn{1}{|>{\columncolor[named]{Green}}r|}{9} &
 \multicolumn{1}{|>{\columncolor[named]{Green}}r|}{10} & \multicolumn{1}{|>{\columncolor[named]{Green}}r|}{11} &
 \multicolumn{1}{|>{\columncolor[named]{Green}}r|}{12} & \multicolumn{1}{|>{\columncolor[named]{Green}}r|}{13} &
 \multicolumn{1}{|>{\columncolor[named]{Green}}r|}{14} & \multicolumn{1}{|>{\columncolor[named]{Green}}r|}{15} \\
\hline

\end{tabular}
\\
\caption[Summary of tests performed on Module 9]{Summary of tests performed on Module 9 (see text).}
\end{table}
\clearpage

%% file: conclusion/conclusion.tex
\section{Conclusions} \label{conclusions}

This document has reported on the tests that were performed to understood and validate the performance of the high voltage system. 

The following points should be noted.
\begin{itemize}
\item{The system should not be operated below 10V.}
\item{If a failure occurs and the system interlock is fired or a current exceeds the maximum value then the voltage is cut suddenly. It is recommended that the effect of this on the VELO modules should be assessed.}
\item{The system can be caused to trip if the system is ramped quickly with a low current limit (e.g ramped at 7V/s and a current limit of $40\mu$A applied).}
\item{As a result of a power supply module firmware error, it is possible to cause instantaneous ramp down of a channel. This occurs if channels in a module are being ramped and another channel is turned on, see section \ref{firmware_error} for details. A temporary protection has been implemented in the PVSS FSM for the use of recipes.}
\item{The accuracy of the current readings of the power supply modules is approximately 1$\mu$A.}
\end{itemize}

The operation of every channel in the system has been tested, the behaviour under various failure modes has been investigated and the noise in the system has been measured. The system is found to meet the requirements of the VELO. 

\section{Acknowledgements} \label{acknowledgements}
We would like to thank Martin Van Beuzekom for his useful suggestions to this document and for conducting the hardware system review of the VELO high voltage system.

%% file: GlasgowPreprint200811_arxiv.bbl
\begin{thebibliography}{99}
\bibitem{HVC} B. Rakotomiaramanana et al., High Voltage Connectivity, EDMS No: 836439
\bibitem{interlock}  N.A. Smith et al., VELO Interlocks, EDMS No: 706629 
\bibitem{HVP} ISEG, High Voltage Power Supply EHQ F607n-F Operator manual, version 1.08, available from  http://njaka.web.cern.ch/njaka/EHQF607n$\_$405-F$\_$V109.pdf
\bibitem{burn-in} A. Bates, F. Marinho et al., VELO module characterisation : Results from the Glasgow LHCb VELO module burn-in, LHCb-2007-103
\bibitem{HVProcedure} VELO High Voltage Procedure, EDMS No: 9096687
\end{thebibliography}
